\documentclass[a4paper,12pt]{article}
 
\pdfoutput=1
\usepackage{jheppub}
\usepackage{dcolumn}
\usepackage{soul}
\usepackage[english]{babel}
\usepackage[utf8]{inputenc}
\usepackage{dcolumn}
\usepackage{soul}
\usepackage{hyperref}
\usepackage{url}
\usepackage{graphicx}
\usepackage{bm,amsmath,amssymb}
\usepackage[mathscr]{eucal}
\usepackage{makeidx}
\usepackage{subfig}
\usepackage{amsmath}
\usepackage{amssymb}
\usepackage{amsthm}
\usepackage{mathrsfs}
\usepackage{graphicx}
\usepackage{fancyhdr}
\usepackage{array}
\usepackage{simplewick}
\usepackage{latexsym}
\usepackage[all]{xy}
\usepackage{enumerate}
\usepackage{dsfont}
\usepackage{slashed}
\usepackage{float}
\usepackage[titletoc]{appendix}
\usepackage{ulem}

\usepackage{verbatim}

\usepackage{tikz,tikz-3dplot}
\tdplotsetmaincoords{80}{45}

\tikzset{surface/.style={draw=black, fill=white, fill opacity=.6}}

\usepackage{tikz}
\usetikzlibrary{decorations.pathmorphing,patterns}

\newcommand{\be}{\begin{equation}}
\newcommand{\ee}{\end{equation}}
\newcommand{\bea}{\begin{eqnarray}}
\newcommand{\eea}{\end{eqnarray}}

\newcommand{\ben}{\begin{eqnarray}}
\newcommand{\een}{\end{eqnarray}}

\title{Building AdS/BCFT Josephson junctions within Horndeski gravity}

\author[a,b,c]{Fabiano F. Santos}\emailAdd{fabiano.ffs23@gmail.com}
\author[d]{Henrique Boschi-Filho}\emailAdd{boschi@if.ufrj.br}
\affiliation[a]{Centro de Ciências Exatas, Naturais e Tecnológicas, UEMASUL, 65901-480, Imperatriz, MA, Brazil.}
\affiliation[b]{School of Physics, Damghan University, Damghan, 3671641167, Iran.}
\affiliation[c]{Departamento de Física, Universidade Federal do Maranhão, São Luís, 65080-805, Brazil.}
\affiliation[d]{Instituto de Física, Universidade Federal do Rio de Janeiro, Rio de Janeiro, 21941-909, Brazil.}

\abstract{This work explores the constriction and normal Josephson junctions of superconductors within Horndeski's gravitational theory framework. Through a single scalar field of this theory, we provide a dual holographic description via the AdS/BCFT correspondence. We identify a critical temperature below which a charged condensate forms through a second-order phase transition in constriction and normal junctions. Our findings reveal that the condensate comprises pairs of quasiparticles. The junctions between superconductors are characterized by weak links that lead to supercurrent flow, with their magnitude determined by the phase difference between the superconductors, which is modulated by the Horndeski parameters. The supercurrents are governed by the Josephson current-phase relation, highlighting the intricate interplay between gravitational theory and superconducting phenomena.} 


\begin{document}
	\maketitle
	\newcommand{\limit}[3]
	{\ensuremath{\lim_{#1 \rightarrow #2} #3}}


\section{Introduction}

Josephson junctions typically consist of two superconductors separated by a weak link \cite{Josephson:1962zz,Hovhannisyan:2022gkp}. The weak link support a phase difference $\Gamma$ between the condensates of the two superconductors which imply a supercurrent flowing across the junction, described by the Josephson current-phase relation:
\begin{equation}
J=J_{max}\sin(\Gamma).  \label{L4} 
\end{equation}
This supercurrent can exist in Josephson junctions without an applied voltage, and the nature of the weak link determines the type of junction. 

Condensed matter physics as superconductors and Josephson junctions are  usually described by electromagnetic interactions, ignoring gravitational and nuclear forces. On the other hand, dualities play an important role in many physical systems. The simplest and possibly the most well known example is the electric-magnetic duality that holds between these fields in vacuum. More sophisticated cases include supersymmetry, as the Seiberg-Witten one \cite{Seiberg:1994rs} and also between different string theories \cite{Witten:1995ex}. One important duality was proposed by Maldacena connecting string theory in 10 dimensional curved spacetime with field theory in flat four dimensions. This is known as the anti-de Sitter/Conformal Field Theory (AdS/CFT) correspondence and relates a string theory on Anti-de Sitter (AdS) spacetimes in the bulk to a Conformal Field Theory (CFT) on its boundary \cite{Maldacena:1997re, Gubser:1998bc, Witten:1998qj, Aharony:1999ti}. This holographic correspondence is a duality between a gravitational theory in higher dimensions and a non-gravitational theory in lower dimensions. It has been applied, for instance, in condensed matter, hydrodynamics, hadronic physics, and the quark-gluon plasma (for reviews, see {\sl e.g.} \cite{Son:2007vk, Erdmenger:2007cm, Schafer:2009dj, Rangamani:2009xk, Casalderrey-Solana:2011dxg, Hartnoll:2016apf, Florkowski:2017olj}). In particular, in Ref. \cite{Hartnoll:2008vx} it was shown how to build a holographic superconductor (see also {\sl e.g.} \cite{Arean:2010xd}), and the application of the AdS/CFT correspondence also allowed the description of holographic Josephson junctions in \cite{Horowitz:2011dz, Wang:2011rva, Wang:2012yj, Domokos:2012rj, Hu:2015dnl}.

A modification of the correspondence, also known as Anti-de Sitter/Boundary Conformal Field Theory (AdS/BCFT), was proposed by Takayanagi \cite{Takayanagi:2011zk},  restriciting the CFT side with an additional boundary. This set up  garnered attention for its potential to investigate transport coefficients, such as those related to the Hall effect, fluid/gravity correspondence, and the Hawking-Page phase transition \cite{Fujita:2011fp, Fujita:2012fp, Melnikov:2012tb, Cavalcanti:2018pta, Magan:2014dwa, Santos:2023flb}. Furthermore, the study of collages of two AdS/BCFT geometries from a common AdS boundary has been explored in the context of defect gravity duals, interface CFT \cite{Erdmenger:2015spo,Bachas:2020yxv,Bachas:2021fqo}, Janus solutions \cite{Aharony:2003qf,Bak:2003jk,DHoker:2007zhm,Bak:2007jm,Azeyanagi:2007qj}, interface junctions \cite{Liu:2024oxg,Liu:2022ezb} and the recently developed double holography \cite{Chen:2020uac,Chen:2020hmv,Geng:2020fxl}. Recently, the AdS/BCFT correspondence also allowed the construction of constriction Josephson junctions \cite{Santos:2024cwf}.

Important extensions of the AdS/CFT correspondence occur when considering alternative gravitational theories impplying corrections to Einstein's equations on the string theory side of the duality. This can be done in a variety of ways, as for instance in the case of scalar-tensor theories. If limited to second order derivatives, the most general formulation of these theories in four dimensions was classified long ago by Horndeski \cite{Horndeski:1974wa}
(for reviews, see, {\sl e.g.} \cite{Koyama:2015vza, Ishak:2018his, Heisenberg:2018vsk, Kobayashi:2019hrl}). 

The inclusion of Horndeski theory in AdS/CFT and in AdS/BCFT correspondences lead to various applications, as for instance, solutions featuring momentum dissipation in dual boundary theories \cite{Jiang:2017imk}. In these models, the chemical potential increases with the Horndeski parameters, which modify the transport properties of the metal/insulator systems \cite{Zhang:2022hxl,Lu:2020ttp}. Notably, increasing Horndeski gravity parameters can hinder phase transitions. As demonstrated in  \cite{Jiang:2017imk}, static electrically charged AdS planar black hole solutions are used to calculate the holographic DC conductivity from the dual field theory \cite{Arean:2010xd,Wang:2011rva,Wang:2012yj,Domokos:2012rj}. These solutions reveal a critical point, where the kinetic term of Horndeski exhibits a ghostly character. This ghostly behavior is linked to nonphysical singular or negative conductivities. The dependence of the conductivity temperature resembles that of a conductor transitioning to semiconductor-like behavior at the critical point. From a transport perspective, the Horndeski gravity framework allows systems to exhibit metallic or insulating characteristics, depending on whether one os its the parameters is negative or positive. Additionally, the thermal conductivities are changed by the intrinsic properties of the black hole and the model parameters \cite{Zhang:2022hxl}. 

The thermodynamics of BTZ black holes within AdS/BCFT and Horndeski gravity has a rich phase structure played by the Horndeski parameters and has been studied in \cite{Santos:2021orr}. A general discussion of the holographic boundary CFT in Horndeski gravity has been presented in \cite{Santos:2024cvx}, and the role of the AdS/BCFT with entanglement with the black hole interior was touched in \cite{Santos:2025wbl}. The complexity growth of the 
black hole entropy in Horndeski gravity within the AdS/BCFT framework was presented in \cite{Santos:2025fdp}.

In this work, we propose the construction AdS/BCFT Josephson junctions within the framework of Horndeski gravity to study the corresponding Josephson current. In our model, the junction of the superconductors is described by the AdS/BCFT duality where we achieve  geometric junctions with two profiles: (i) a constriction-like junction;  and (ii)  a Superconductor/Normal/Superconductor (SNS) junction. They are derived from the action as field equations for the extrinsic curvature, where the weak link is represented by a  scalar field.




\section{Summary of main results and novelty}

\begin{itemize}
\item First, we construct, for the first time, Josephson junctions in the AdS/BCFT framework within Horndeski gravity, where the junction geometry is controlled by the profile of the end‑of‑the‑world brane and by the Horndeski couplings.
\item Second, we show that both constriction‑type and normal SNS‑type junctions can be realized in a unified setup, and we derive analytic expressions for the condensate and for the maximum Josephson current in terms of the Horndeski parameters,
\item Third, we demonstrate how the Horndeski couplings deform the Josephson relation and the coherence length, providing a controlled way to interpolate between different effective junction geometries from the gravitational side.
\item These features go beyond previous works on holographic Josephson junctions in Einstein gravity and earlier AdS/BCFT–Horndeski constructions, and we believe they are of independent interest for the holographic modeling of superconducting interfaces.
\end{itemize}


This work is organized as follows. In Sect. \ref{JJH}, we present the general ste up for our proposals for the Josephson junctions constructed within Horndeski gravity and using the AdS/BCFT correspondence. In Sect.  \ref{Hornd}, we present the equations governing the holographic setting within the Horndeski gravity, and in Sect. \ref{black}, the corresponding black hole solutions using the probe approximation. In Sect. \ref{prof1}, we present a detailed formulation of the constriction Josephson junctions in Horndeski gravity following some profile solutions of Horndeski gravity and deduce the corresponding Josephson current. In Sect. \ref{prof2}, we present the profile of the normal Josephson junctions in Horndeski gravity together with the resulting current. Finally, we present our conclusions and discussions in Sect. \ref{concl}. 

\section{Proposal of Josephson junctions within AdS/BCFT and Hornkeski gravity} \label{JJH}

\begin{figure}[!ht]
\begin{center}
\includegraphics[width=\textwidth]{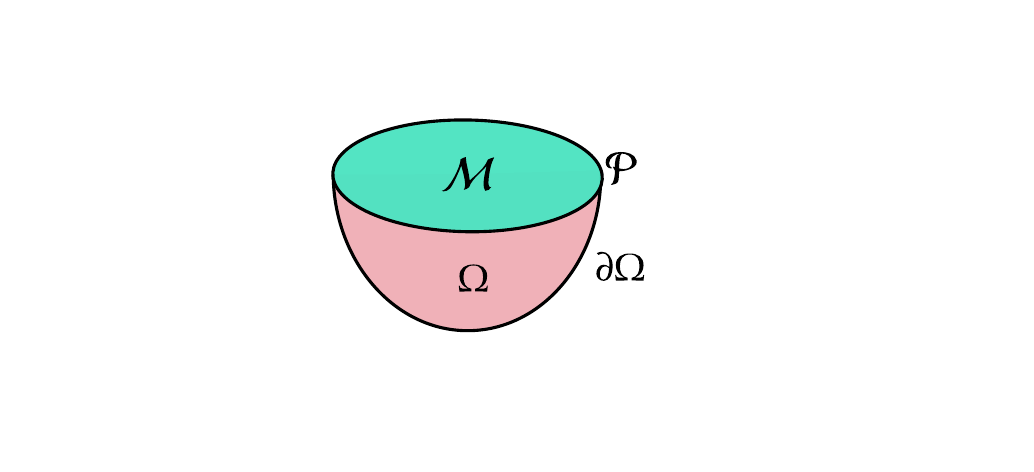}
\vskip -1.5cm
\caption{AdS/BCFT correspondence in the presence of boundary hypersurface $\Omega$.}\label{p}
\end{center}
\end{figure}

The AdS/BCFT correspondence, as proposed by Takayanagi \cite{Takayanagi:2011zk}, modifies the usual AdS/CFT duality  \cite{Maldacena:1997re} by defining a boundary $\mathcal{M}$ for a $(d+1)$ dimensional asymptotically AdS space \cite{Santos:2024cvx,Santos:2025wbl,Santos:2025fdp,Geng:2023iqd,Geng:2020qvw,Geng:2021wcq,Geng:2021iyq,Geng:2021mic,Geng:2022dua,Geng:2023qwm,Geng:2022slq,Geng:2022tfc,Geng:2024xpj,Geng:2025efs,Bao:2025plr,Geng:2025yys,Paul:2025gpk}, such that $\Omega=\mathcal{M}\,\cup\,\partial\Omega$, where  $\partial\Omega$ is a $d$ dimensional manifold that satisfies $\partial\Omega\,\cap\,\mathcal{M}=\mathcal{P}$ (see Fig. \ref{p}).

\begin{figure}[!ht]
\begin{center}
\includegraphics[width=\textwidth]{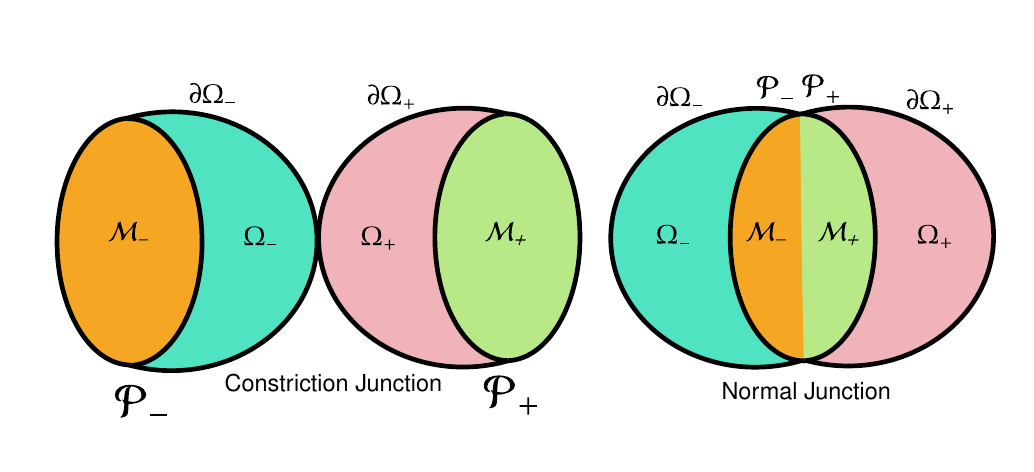}
\caption{The figure presents two possible configurations for the bulk manifold into the two domains $\Omega_{\pm}$ that can describe Josephson junctions. {\sl Left}: In the AdS/BCFT correspondence scheme $\mathcal{M}_{\pm}$ are asymptotically AdS spaces so that $\Omega_{\pm}=\mathcal{M}_{\pm}\cup\partial\Omega_{\pm}$ where $\partial\Omega_{\pm}$ are  manifolds that satisfy $\partial\Omega_{\pm}=\mathcal{M}_{\pm}=\mathcal{P}_{\pm}$. {\sl Right}: we  glued two Anti-de Sitter (AdS) geometries \cite{Kawamoto:2023wzj,Kanda:2023zse} that form the normal junction. In this case, one couples two brane field theories (BFTs) via gravitational interactions in the dual holographic perspective that corresponds to the case $\mathcal{M}_{\pm}=\mathcal{P}_{\pm}$ (where in these regions the scalar field represents the brane).}
\label{p1}
\end{center}
\end{figure}

 A crucial component in constructing the gravitational dual of a Josephson junction is the phase difference \cite{Hu:2015dnl,Horowitz:2011dz,Hartnoll:2008vx}. We propose that this phase difference can be expressed as $\Psi=\phi\,e^{i\theta}$, where $\Psi$ is a complex charged scalar field, $\phi$ is a real scalar field, and $\theta$ is the Stückelberg field \cite{Liu:2022bdu}. The Lagrangian representing these fields together with the electromagnetic field strength $F=dA$, providing a comprehensive framework for analyzing the Josephson junction within this holographic setup is given by
\begin{eqnarray}
&&{\cal L}_{\rm FF}=-\dfrac{\kappa}{4} F^{\mu \nu} F_{\mu \nu}-(\partial_\mu\Psi-qA_\mu)^2-\frac{1}{2}m^2\Psi^2, \label{L3} 
\end{eqnarray}
where $m$ is the mass of the scalar field $\Psi$. We restrict our study to the probe approximation scenario, where the backreaction of the fields on the metric is negligible. To achieve this, we perform a rescaling of the fields defined as $\Psi=\tilde{\Psi}/q$ and $A=\tilde{A}/q$, and take the limit $q\to\infty$ while keeping $\tilde{\Psi}$ and $\tilde{A}$ fixed. This approach ensures that the influence of the fields on the spacetime geometry can be ignored, allowing us to focus on the dynamics of $\tilde{\Psi}$ and $\tilde{A}$ without considering their gravitational effects.
Holographic superconductors within the AdS/CFT correspondence typically employ Dirichlet boundary conditions at the AdS boundary; these conditions are applied in $\mathcal{M}$. However, for the holographic Josephson junction in the AdS/BCFT framework, as discussed in \cite{Takayanagi:2011zk,Fujita:2011fp,Santos:2024cwf}, a Neumann boundary condition (NBC) on $\partial\Omega$ is required to account for the dynamical nature of this boundary. The absence of a naturally defined metric on $\Omega$ on the CFT side requires this approach \cite{Nozaki:2012qd}. This duality is rooted in the holographic derivation of entanglement entropy \cite{Ryu:2006bv,DosSantos:2022exb} and the Randall-Sundrum model \cite{Randall:1999vf}. The extension of the CFT boundary to the AdS bulk is similar to a modification of a thin Randall-Sundrum brane intersecting the AdS boundary, which must be dynamic and subject to NBC \cite{Takayanagi:2011zk,Fujita:2011fp}. In this framework, any discontinuity in the bulk extrinsic curvature across the defect is balanced by the brane tension \cite{Santos:2021orr,Santos:2023flb}.

In our study, the Neumann Boundary Condition (NBC) specifies the boundary $\mathcal{P}$ by the condition $y=const$, where $y$ is a coordinate on
$\mathcal{M}$ (see Fig. \ref{p}). This AdS/BCFT scenario corresponds to a problem defined in half of Minkowski space \cite{Santos:2021orr,Santos:2023flb}. As demonstrated by \cite{Santos:2023flb}, the Dirichlet Boundary Condition (DBC) $n^{\mu}M_{\mu\nu}|_{\partial\Omega}=0$ provides a profile with corrections from the Horndeski parameters. We derive the transport coefficients using NBC within Horndeski gravity and analyze constriction and normal Josephson junctions. At the normal Josephson junction, we examine the evolution of the Horndeski $\gamma$ parameter, where the weak link can exhibit superconductor/normal/superconductor (SNS) behavior \cite{Santos:2024cwf,Siano,Chandrasekhar:2024epu,Senapati,A.Baselmans,Horowitz:2011dz}. We discuss the role of $J_{max}$ in our model and its implications for the size of the weak link as we vary the Horndeski gravity parameters. This model is motivated by the potential to extend the range of condensed matter phenomena that can be described using gravitational analogs \cite{Horowitz:2011dz}.

A Josephson junction-like constriction is created using two profiles that form the constriction junction \cite{Santos:2024cwf,Siano,Chandrasekhar:2024epu,Senapati} (see the left side of Fig. \ref{p1}). The Superconductor/Normal/Superconductor (SNS) junctions are derived from the action as field equations for the extrinsic curvature, where the weak link is represented by the scalar field $\phi=\phi_{+}\cup\,\phi_{-}$ between $\partial\Omega=\partial\Omega_{+}\cup\,\partial\Omega_{-}$
\cite{A.Baselmans,Horowitz:2011dz,Erdmenger:2014xya,Minamitsuji:2013vra,Kawamoto:2023wzj,Kanda:2023zse} (see the right side of Fig. \ref{p1}).

\section{Holographic framework in Horndeski gravity}\label{Hornd}

In this section, we construct a holographic model to study Josephson junctions within Horndeski gravity. In this scenario, the AdS/BCFT setup requires supplementing the bulk action with an appropriate generalized Gibbons–Hawking–York boundary term so that the variational problem is well posed on the hypersurface $\partial\Omega$. In addition, in asymptotically AdS spacetimes one introduces boundary counterterms within the holographic renormalization scheme in order to render the on‑shell action and the associated conserved quantities finite \cite{Santos:2021orr,Santos:2023flb}. These ingredients underlie the AdS/BCFT construction that we employ in the present work. 
This will be described through the following total action.
\begin{eqnarray}
S&=&S^{\Omega}+S^{\Omega}_{mat}+S^{\partial\Omega}+S^{\partial\Omega}_{mat}+S^{\mathcal{P}}_{ct}+S^{\Omega}_{FF}\,, \label{HF1}
\end{eqnarray}
where $S^{\mathcal{M}}_{mat}$ describes ordinary matter that is supposed to be a perfect fluid, and 
\begin{eqnarray}
&&S^{\Omega}=\kappa\left(\int_{\Omega_{+}}+\int_{\Omega_{-}}\right){d^{4}x\sqrt{-g}}{\Big[R-2\Lambda-\frac{1}{2}(\alpha g_{\mu\nu}-\gamma\,  G_{\mu\nu})\nabla^{\mu}\phi\nabla^{\nu}\phi\Big]},
\end{eqnarray}
\begin{eqnarray}
&&S^{\Omega}_{FF}=\frac{\kappa}{q^{2}}\left(\int_{\Omega_{+}}+\int_{\Omega_{-}}\right){d^{4}x\sqrt{-g}}{\Big[-\dfrac{1}{4} F^{\mu \nu} F_{\mu \nu}-|\partial_\mu\Psi-qA_\mu|^2-\frac{1}{2}m^2|\Psi|^2\Big]},\label{EMAX}
\end{eqnarray}
\begin{eqnarray}
&&S^{\partial\Omega}=2\kappa\left(\int_{\partial\Omega_{+}}+\int_{\partial\Omega_{-}}\right)
d^{3}x\sqrt{-h}\Big\{K-\Sigma-\frac{\gamma}{4}(\nabla_{\mu} \phi\nabla_{\nu} \phi)K^{\mu\nu}
\nonumber\\ 
&&\hskip 6cm -\frac{\gamma}{4}\big[(\nabla_{\mu}\phi\nabla_{\nu}\phi\,n^{\mu}n^{\nu}+(\nabla \phi)^2)K  \big]\Big\},\label{SdO}
\end{eqnarray}
\begin{eqnarray}
&&S^{\partial\Omega}_{mat}=2\left(\int_{\partial\Omega_{+}}+\int_{\partial\Omega_{-}}\right){d^{3}x\sqrt{-h}\mathcal{L}_{mat}},
\end{eqnarray}
\begin{eqnarray}
&&S^{\mathcal{P}}_{ct}=2\kappa\left(\int_{\mathcal{P}_{+}}+\int_{\mathcal{P}_{-}}\right){d^{3}x\sqrt{-h}(c_{0}+c_{1}R+c_{2}R^{ij}R_{ij}+c_{3}R^{2}+b_{1}(\partial_{i}\phi\partial^{i}\phi)^{2}+\cdots)}\,. \nonumber\\
\end{eqnarray}
The solution presented in Equation (\ref{EMAX}) resembles the case discussed in \cite{Horowitz:2011dz}, with a key distinction arising from the Horndeski parameters $\alpha$ and $\gamma$ in the function $f(u)$. The Horndeski gravity theory, with a single scalar field, is described through its Lagrangian: 
\begin{eqnarray}\label{eq:Lhorn}
{\cal L}_{\rm H}=\kappa(R-2\Lambda)-\frac{1}{2}(\alpha g_{\mu\nu}-\gamma\,  G_{\mu\nu})\nabla^{\mu}\phi\nabla^{\nu}\phi,\label{L1}
\end{eqnarray}
where $R$, $G_{\mu \nu}$, and $\Lambda$ are the scalar curvature, the Einstein tensor, and the cosmological constant, respectively; $\phi$ is a scalar field, $\alpha$ and $\gamma$ are coupling constants, while $\kappa={1}/{(16 \pi G_N)}$, with $G_N$ being the Newtonian gravitational constant. The Lagrangian $\mathcal{L}_{bdry}$ 
defined on the the boundary action $S^{\partial\Omega}$, Eq. \eqref{SdO}, has already been proposed in \cite{Santos:2024cvx,Santos:2025wbl,Santos:2025fdp} as
\begin{equation}
\mathcal{L}_{bdry}=(K-\Sigma)-\frac{\gamma}{4}(\nabla_{\mu}\phi\nabla_{\nu}\phi n^{\mu}n^{\nu}-(\nabla \phi)^2)K-\frac{\gamma}{4}\nabla_{\mu}\phi\nabla_{\nu}\phi K^{\mu\nu}\,. 
\end{equation}
Here $K_{\mu\nu}=h^{\phantom{\mu}\beta}_{\mu}\nabla_{\beta}n_{\nu}$ is the extrinsic curvature, $K=h^{\mu\nu}K_{\mu\nu}$ is the trace of the extrinsic curvature, $h_{\mu\nu}$ is the induced metric, $n^{\mu}$ is an outward pointing unit normal vector to the hypersurface boundary $\partial\Omega$, and $\Sigma$ is the boundary tension on $\partial\Omega$. On the other hand, $\mathcal{L}_{mat}$ is the Lagrangian matter in $\Omega$. Since we provide an asymptotic AdS spacetime, we need to impose the boundary counterterms ${\cal L}_{ct}$ \cite{Santos:2024cvx,Santos:2025wbl,Santos:2025fdp}, which are given by
\begin{equation}
\mathcal{L}_{ct}=c_{0}+c_{1}R+c_{2}R^{ij}R_{ij}+c_{3}R^{2}+b_{1}(\partial_{i}\phi\partial^{i}\phi)^{2}+\cdots.
\end{equation}
These do not affect the bulk dynamics and will be neglected. From the action (\ref{HF1}) and considering the scaling $\Psi=\tilde{\Psi}/q$, $A=\tilde{A}/q$ with $q\to\infty$, we derive the equations of motion as follows.
\begin{eqnarray}
&&{\cal E}_{\mu\nu}[g_{\mu\nu},\phi]_{\Omega}=G_{\mu\nu}+\Lambda g_{\mu\nu}-{\cal T}_{\mu\nu},\\
&&{\cal E}^{\Omega}_{\phi}=\nabla_{\mu}\left[\left(\alpha g^{\mu\nu}-\gamma G^{\mu\nu}\right)\nabla_{\nu}\phi\right]\,,\label{H5}\\
&&{\cal F}^{\partial\Omega}_{\phi}=-\frac{\gamma}{4}(\nabla_{\mu}\nabla_{\nu}\phi n^{\mu}n^{\nu}-(\nabla^{2}\phi))K-\frac{\gamma}{4}(\nabla_{\mu}\nabla_{\nu}\phi)K^{\mu\nu}\,,\label{H6}\\
&&[\nabla_{\mu}F^{\mu\nu}-2q^{2}A^{\mu}-m^2|\Psi|]_{\Omega}=0,\label{Eq:M}
\end{eqnarray}
with
\begin{eqnarray}
{\cal T}_{\mu\nu}&=&\frac{\alpha}{2}\left(\nabla_{\mu}\phi\nabla_{\nu}\phi-\frac{1}{2}g_{\mu\nu}\nabla_{\lambda}\phi\nabla^{\lambda}\phi\right)\label{11}\nonumber\\
                  &+&\frac{\gamma}{2}\left(\frac{1}{2}\nabla_{\mu}\phi\nabla_{\nu}\phi R-2\nabla_{\lambda}\phi\nabla_{(\mu}\phi R^{\lambda}_{\nu)}-\nabla^{\lambda}\phi\nabla^{\rho}\phi R_{\mu\lambda\nu\rho}\right)\nonumber\\
									&-&\frac{\gamma}{2}\left(-(\nabla_{\mu}\nabla^{\lambda}\phi)(\nabla_{\nu}\nabla_{\lambda}\phi)+(\nabla_{\mu}\nabla_{\nu}\phi)\Box\phi+\frac{1}{2}G_{\mu\nu}(\nabla\phi)^{2}\right)\nonumber\\
									&-&\frac{\gamma\,g_{\mu\nu}}{2}\left(-\frac{1}{2}(\nabla^{\lambda}\nabla^{\rho}\phi)(\nabla_{\lambda}\nabla_{\rho}\phi)+\frac{1}{2}(\Box\phi)^{2}-(\nabla_{\lambda}\phi\nabla_{\rho}\phi)R^{\lambda\rho}\right),
\end{eqnarray}
where $\Omega=\Omega_{+}\cup\,\Omega_{-}$, $\partial\Omega=\partial\Omega_{+}\cup\,\partial\Omega_{-}$ and note that ${\cal E}^{\Omega}_{\phi}={\cal F}^{\partial\Omega}_{\phi}$, from the Euler-Lagrange equation \cite{Santos:2021orr,Santos:2023flb,Santos:2024cvx,Santos:2025wbl,Santos:2025fdp}. ${\cal S}^{\partial\Omega}_{\mu\nu}$ is the boundary tensor located at hypersurface $\partial\Omega$, which will be presented in the section \ref{prof1}.

\section{AdS planar Schwarzschild black hole}\label{black}
 In this section, we present a fixed metric background AdS planar Schwarzschild black hole.
 \begin{equation}
 ds^{2}=-f(u)dt^2+\frac{du^2}{f(u)}+u^2(dx^2+dy^2).\label{metric}
 \end{equation}
As discussed by \cite{Santos:2021orr,Santos:2023flb,Santos:2024cvx,Santos:2025wbl,Santos:2025fdp,DosSantos:2022exb}, in order to escape from the no-hair theorem \cite{Rinaldi:2012vy,Babichev:2013cya,Anabalon:2013oea}, we need to impose that ${\cal E}_{\phi}[g_{rr},\phi]={\cal E}_{rr}[g_{rr},\phi]=0$, which provides 
\begin{equation}
\alpha g_{uu}-\gamma G_{uu}=0\label{hair1}\,.
\end{equation}
This condition is satisfied through gauge fixation, performed for static black hole configurations to contour the no-hair theorems \cite{Hui:2012qt}. Specifically, the square of the radial component of the conserved current must vanish identically without imposing restrictions on the radial dependence of the scalar field, as indicated in equation (\ref{hair1}). Our study focuses on holographic models of boundary CFTs at finite temperatures. Within this framework, the solution to the AdS/BCFT problem is characterized by an asymptotically AdS$_{4}$ black hole geometry. The condition outlined in Equation (\ref{hair1}) provides the basis for this solution, which is expressed as
\begin{eqnarray}
&&f(u)=\frac{\alpha}{3\gamma}\left(u^{2}-\frac{u^{3}_{h}}{u}\right),\label{sol1}\\
&&\psi^{2}(u)=-\frac{2\kappa(\alpha+\gamma\Lambda)}{\alpha\gamma}\frac{1}{f(u)}.\label{sol2}
\end{eqnarray}
In this context, $u_{h}$ denotes the black hole horizon, and $\phi{'}=\psi$. The solution presented in equation (\ref{sol1}) resembles the case discussed in \cite{Horowitz:2011dz}, with a key distinction arising from the Horndeski parameters $\alpha$ and $\gamma$ in the function $f(u)$. These parameters introduce significant modifications to the properties of black holes. The temperature of this black hole can be expressed as
\begin{eqnarray}
T=\frac{\alpha}{\gamma}\frac{u_{h}}{4\pi}.\label{sol3}
\end{eqnarray}
Now, we introduce an additional component for constructing the constriction Josephson junction in Horndeski gravity, as illustrated in Fig. \ref{p1}, oriented along the $x$ direction with fields that are functions of both $u$ and $x$. The phase difference in the Josephson junction requires the incorporation of a phase into our scalar field, expressed as $\Psi=\phi\,e^{i\theta}$. To solve the equation (\ref{Eq:M}), we consider the following solutions:
\begin{eqnarray}
\tilde{A}=A_{t}dt+A_{u}du+A_{x}dx,\label{sol4}
\end{eqnarray}
where $\Psi$, $A_{t}$, $A_{u}$, and $A_{x}$ are all real functions of $u$ and $y$. In addition to the gauge field $A$, there are gauge-invariant fields $M=A-d\theta$ \cite{Santos:2024cwf,Horowitz:2011dz}. Now, with these gauge transformations, we derive the equations of motion in terms of gauge-invariant quantities as follows.
\begin{eqnarray}
&\partial^{2}_{u}|\Psi|+\frac{}{u^2f}\partial^{2}_{x}|\Psi|+\left(\frac{f^{'}}{f}+\frac{2}{u}\right)\partial_{u}|\Psi|+\frac{1}{f}\left(\frac{M^{2}_{t}}{f}-fM^{2}_{u}-\frac{M^{2}_{x}}{u^2}-m^2\right)|\Psi|=0,\label{a}\\
&\partial^{2}_{u}M_{t}+\frac{1}{u^{2}f}\partial^{2}_{x}M_{t}+\frac{2}{u}\partial_{u}M_{t}-\frac{2|\Psi|^{2}}{f}M_{t}=0,\label{b}\\
&\partial^{2}_{u}M_{u}-\partial_u\partial_xM_x-2u^2|\Psi|^2M_u=0,\label{c}\\
&\partial^{2}_{u}M_{x}-\partial_u\partial_xM_u+\frac{f^{'}}{f}(\partial_uM_x-\partial_xM_u)-\frac{2|\Psi|^2}{f}M_x=0,\label{d}\\
&\partial_uM_u+\frac{1}{u^2f}\partial_xM_x+\frac{2}{|\Psi|}\left(M_u\partial_u|\Psi|+\frac{M_x}{u^2f}\partial_x|\Psi|\right)+\left(\frac{f^{'}}{f}+\frac{2}{u}\right)M_u=0.\label{e}
\end{eqnarray}
The above equations (\ref{a})-(\ref{d}) are second-order dynamical equations, while the last one (\ref{e}) is a first-order constraint equation arising from the conservation of the source in Maxwell's equations \cite{Santos:2024cwf,Horowitz:2011dz}. Consequently, we can express the asymptotic form of the Maxwell fields as follows.
\begin{eqnarray}
&&M_t=\mu-\frac{\rho(y)}{u}+\mathcal{O}\left(\frac{1}{u^{2}}\right),\\
&&M_u=\mathcal{O}\left(\frac{1}{u^{3}}\right),\\
&&M_y=\nu(y)+\frac{J}{u}.
\end{eqnarray}
The above quantities $\mu$, $\rho(y)$, $\nu(y)$, and $J$ are interpreted in the boundary field theory as chemical potential, charge density, superfluid velocity, and current, respectively \cite{Hu:2015dnl}. However, in the boundary conditions of $u\to\infty$ (with the values $L_{AdS}=1$ and $m=-2$), the scalar field exhibits an asymptotic behavior similar to the Horndeski parameters:
\begin{eqnarray}
&&|\Psi|=\frac{\sqrt{|-\xi|}\Psi^{(1)}(y)}{u}+\frac{\sqrt{|-\xi|}\Psi^{(2)}(y)}{u^{2}}+\mathcal{O}\left(\frac{1}{u^{3}}\right),\label{Psi} 
\end{eqnarray}
where we defined 
\begin{eqnarray}
&&\xi=\frac{2(\alpha+\gamma\Lambda)}{\alpha^2},  
\label{xi}
\end{eqnarray}
as a real parameter for any values of $\alpha$ and $\gamma$.

In our study, we introduce new bulk fields relative to \cite{Santos:2024cwf,Horowitz:2011dz}. For the choice $m^2 = -2$ and $L_{AdS} = 1$, the two independent fall‑off modes $\Psi^{(1)}(y)$ and $\Psi^{(2)}(y)$ are both normalizable, and one may work either in the standard or in the alternative quantization. Throughout this work we adopt the standard quantization, in which $\Psi^{(1)}(y)$ plays the role of the source for a dimension‑two operator $\mathcal{O}$ in the boundary theory while $\Psi^{(2)}(y)$ is proportional to its expectation value, $\langle \mathcal{O}(y) \rangle = \Psi^{(2)}(y)$ \cite{Baggioli:2016rdj,Son:2002sd,Hartnoll:2009sz,Jiang:2017imk,Zhang:2022hxl,Lu:2020ttp}. In the Josephson‑junction configurations of interest we set the explicit source to zero at the boundary, $\Psi^{(1)}(y) = 0$, so that the condensate forms spontaneously. The corresponding bulk scalar field $\Psi$ is dual to the operator $\mathcal{O}$ and couples to it through a boundary deformation of the form 
\begin{eqnarray}
\mathcal{L}_{CFT}+\int{d^{d}x\varphi_0\mathcal{O}}
\end{eqnarray}
and through Quantum Field Theory (QFT) \cite{Baggioli:2016rdj,Hartnoll:2009sz,Jiang:2017imk} the functional generator of correlation functions $W(\varphi_0)$ can be written as
\begin{eqnarray}
e^{W(\varphi_0)}\langle\,e^{\int{\varphi_0\mathcal{O}}}\rangle_{QFT}. 
\end{eqnarray}
Applying the methodology of \cite{Jiang:2017imk,Zhang:2022hxl,Josephson:1962zz,Lu:2020ttp}, we can derive the correlation functions of the operator $\mathcal{O}$, taking functional derivatives of the generating functional with respect to the source term $\varphi_0$:
\begin{eqnarray}
(\langle \mathcal{O}{...}\mathcal{O} \rangle)_{1...n}=\frac{\delta^{n}W}{\delta\varphi^{n}_{0}}\mid_{\varphi_{0}=0}. 
\end{eqnarray}
 This process allows us to systematically explore the boundary theory's response to perturbations in the bulk. This provides the relationship between bulk fields and boundary operators. Computing these correlation functions gives expectation values for our model. 

\section{Profile description of the constriction Josephson junctions}\label{prof1}

\begin{figure}[!ht]
\begin{center}
\includegraphics[width=\textwidth]{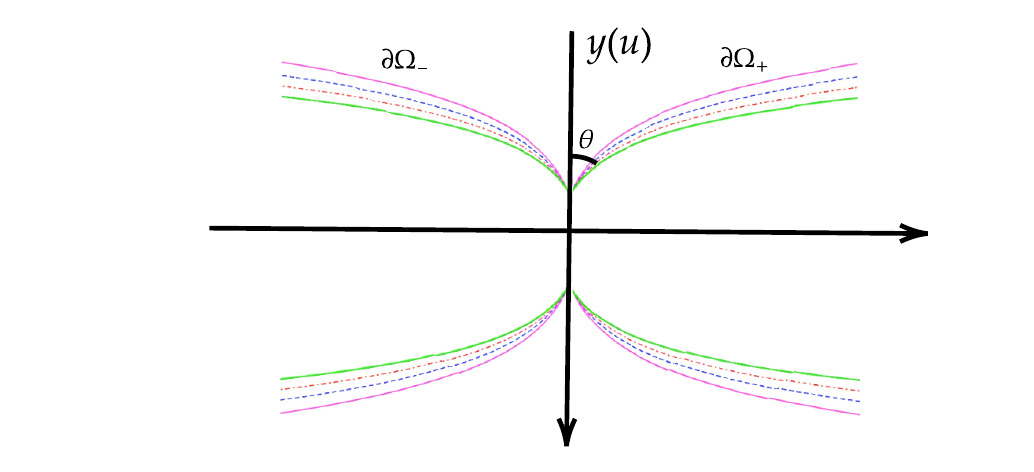}
\caption{The figure presents the junction of two profiles with the two domains $\partial\Omega_{\pm}$ for the values $\theta=\theta'-2\pi/3$, $\Sigma=1/4$, $\Lambda=-1$, $\alpha=8/3$ with $\gamma=0$ ({\sl solid}), $\gamma=-0.1$ ({\sl dashed}), $\gamma=-0.2$ ({\sl dot dashed}), and $\gamma=-0.3$ ({\sl thick}).}
\label{figpro}
\end{center}
\end{figure}

Now, we can alternatively choose the Neumann boundary conditions. Let us do this for the induced metric on $\partial\Omega$, that is, let us impose \cite{Santos:2021orr,Santos:2023flb}:
\begin{eqnarray}
\left[K_{\mu\nu}-h_{\mu\nu}(K-\Sigma)-\frac{\gamma}{4}H_{\mu\nu}\right]_{\partial\Omega}=\kappa {\cal S}^{\partial\Omega}_{\mu\nu}\,,\label{H7}
\end{eqnarray}
where
\begin{eqnarray}
&&{H_{\alpha\beta}\equiv(\nabla_{\mu}\phi\nabla_{\nu}\phi \, n^{\mu}n^{\nu}-(\nabla\phi)^{2})(K_{\alpha\beta}-h_{\alpha\beta}K)-(\nabla_{\mu}\phi\nabla^{\mu}\phi)h_{\alpha\beta}K}\,,\label{H8}\\
&&{\cal S}^{\partial\Omega}_{\alpha\beta}=-\frac{2}{\sqrt{-h}}\frac{\delta S^{\partial\Omega}_{mat}}{\delta h^{\alpha\beta}}\,.\label{H9} 
\end{eqnarray}
Taking into account the stress-energy tensor of matter in $\partial\Omega$ as a constant (this is ${\cal S}^{\partial\Omega}_{\alpha\beta}=0$), we can write
\begin{eqnarray}
\left[K_{\mu\nu}-h_{\mu\nu}(K-\Sigma)-\frac{\gamma}{4}H_{\mu\nu}\right]_{\partial\Omega}=0\,.\label{H10}
\end{eqnarray}
This is the dynamical equation for the induced metric $h_{\mu\nu}$, which describes the hypersurface profile $\partial\Omega$. Solving the AdS/BCFT problem requires finding solutions to the equations of motion derived from the action (\ref{HF1}). These solutions must satisfy the Dirichlet boundary conditions in $\mathcal{M}$ and the Neumann boundary conditions (\ref{H10}) in $\partial\Omega$. Using the metric (\ref{metric}), we can construct the induced metric using the following normal vector:
\begin{eqnarray}
(n^{t},n^{x},n^{y},n^{u})=\left(0,0,\frac{1}{ug(u)},-\frac{uf(u)y{'}(u)}{g(u)}\right),\label{4}
\end{eqnarray}
where these normal vectors are given on the surface $\partial\Omega$ with $g^{2}(z)=1+u^2f(u)y'^{2}(u)$ and $y{'}={dy}/{du}$, and with this, the induced metric is given by
\begin{eqnarray}
 ds^{2}_{ind}=-f(u)dt^2+\frac{g^2(u)du^2}{f(u)}+u^2dx^2.\label{metric1}
\end{eqnarray}
Now, solving equation (\ref{H10}) for the induced metric on the boundary $\partial\Omega$, we have
\begin{eqnarray}
y{'}_{\pm}(u)=\pm\frac{\Sigma}{\sqrt{4+\dfrac{\gamma\,\psi^{2}(u)}{4}+\Sigma^{2}u^2f(u)}},\label{prof}
\end{eqnarray}
in terms of $f(u)$ and $\psi^{2}(u)$, given by Eqs. \eqref{sol1} and \eqref{sol2}, respectively. 
The behavior of the above equation is shown in Fig. \ref{figpro}. Our Josephson junctions are composed of two profiles that form the constriction junction.

\begin{figure}[!ht]
\begin{center}
\includegraphics[width=\textwidth]{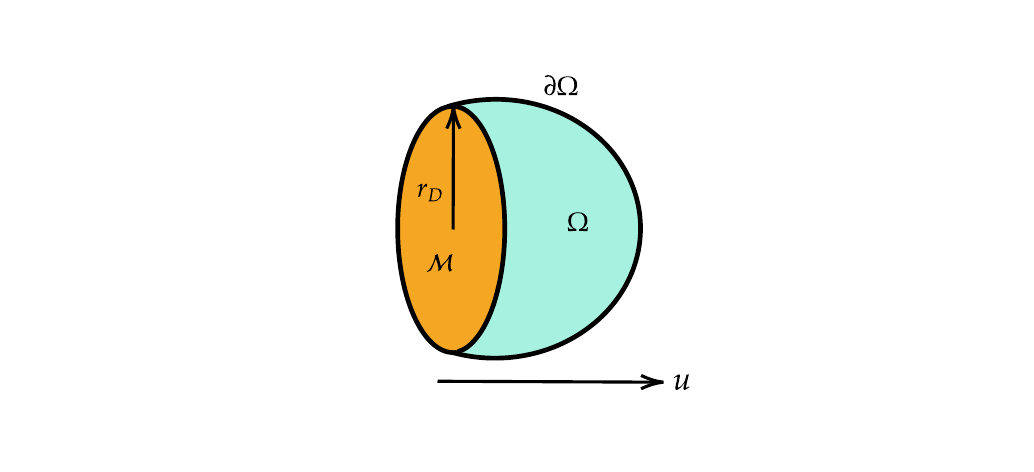}
\caption{The holographic dual of a disk.}
\label{DISK}
\end{center}
\end{figure}

Following the steps of \cite{Santos:2024cwf, Horowitz:2011dz}, the phase difference between the condensate of two superconductors has contributions from Horndeski gravity through the boundary profile is given by gauge invariance:
\begin{eqnarray}
\Gamma=\Delta\theta-\int{A_{y}},
\end{eqnarray}
where the integral crosses the gap. The edges of the gap will not be completely sharp (for numerical reasons). Analogously \cite{Santos:2024cwf}, we have the following:
\begin{eqnarray}
\Gamma=-\frac{1}{J}\int^{+\infty}_{-\infty}{dy[\nu(y)-\nu(\pm\infty)-\mathcal{O}{\rm (complex\,\,\,terms)}]},\label{phase1}
\end{eqnarray}
In this new prescription, $y(u)$ plays the role of $\mu(x)$, as present in \cite{Horowitz:2011dz}, where the weak link between the constriction junction is supported by the boundary tension $\Sigma$, which is how the Horndeski parameters control the Josephson current. Furthermore, Josephson's current, which satisfies
\begin{eqnarray}
J=J_{max}\sin(\Gamma),\label{currentjsp}
\end{eqnarray}
exists without any applied voltage. Some types of Josephson junctions depend on the nature of the link \cite{Santos:2024cwf,Siano,Chandrasekhar:2024epu,Senapati,A.Baselmans,Horowitz:2011dz}. But here, the nature of the link is related to the tension $\Sigma$. Using the Dirichlet Boundary Conditions $n^{\mu}M_{\mu}|_{\partial\Omega}=0$ \cite{Santos:2024cwf}, we have
\begin{eqnarray}
\nu(y)=\frac{J}{u},
\end{eqnarray}
where $y=\int{du\,y{'}(u)}$. Recalling equation (\ref{phase1}) with $\nu(y)|_{y\to\infty}\to\infty$, we solve this equation using the black hole solution $f(u)$. Our Josephson current flows from the disk ($\tau^2+y^2\leq\,r_{D}$ Fig. \ref{DISK}, $\tau$ is Euclidean time) on the surface $\mathcal{M}$ to the boundary at $\partial\Omega$ \cite{Takayanagi:2011zk,Santos:2024cwf,Fujita:2011fp}. So, we have
\begin{eqnarray}
\Gamma&=&-\int^{+\infty}_{-\infty}{du\frac{y{'}(u)}{u}}\nonumber\\
&=&-\int^{-u_h}_{-\infty}{\frac{du}{u}\frac{\Sigma}{\sqrt{4+\dfrac{\gamma\psi^{2}(u)}{4}+\Sigma^{2}u^2f(u)}}}
\nonumber\\
&&+\int^{\infty}_{u_h}{\frac{du}{u}\frac{\Sigma}{\sqrt{4+\dfrac{\gamma\psi^{2}(u)}{4}+\Sigma^{2}u^2f(u)}}}\,. 
\qquad 
\label{phase2}
\end{eqnarray}

\begin{figure}[!ht]
\begin{center}
\includegraphics[scale=0.5]{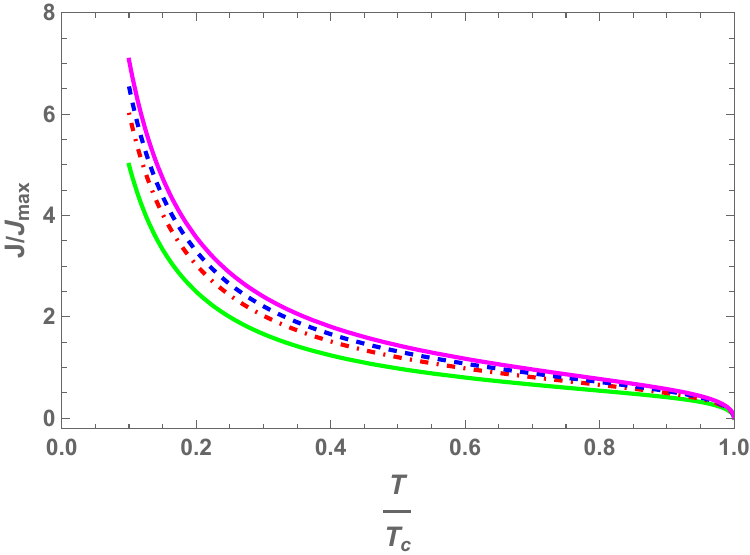}
\includegraphics[scale=0.5]{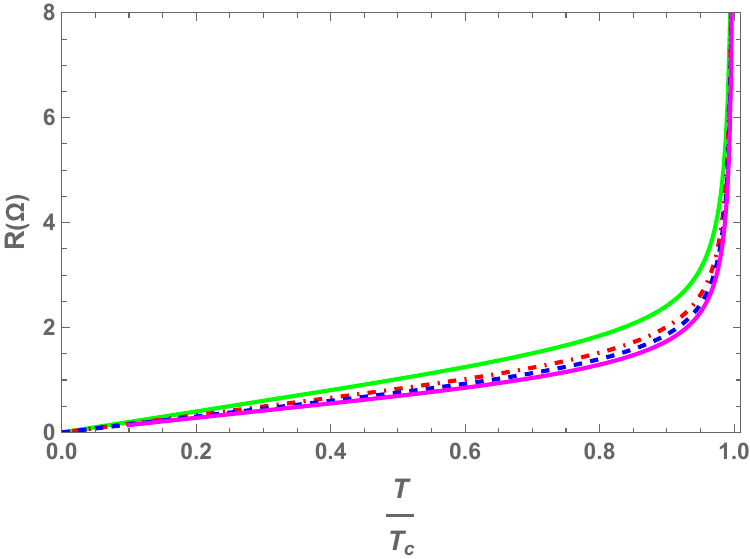}
\caption{The figure shows the behavior of the equation (\ref{currentjsp}). In the left side for the values $\theta=\theta'-2\pi/3$, $\Lambda=-1$, $\alpha=8/3$, $\Sigma=1/4$ with $\gamma=-0.1$ ({\sl solid}), $\gamma=-0.2$ ({\sl dashed}), $\gamma=-0.3$ ({\sl dot dashed}), and $\gamma=-0.4$ ({\sl thick}). In the right side for the values $\gamma=0.1$ ({\sl solid}), $\gamma=0.2$ ({\sl dashed}), $\gamma=0.3$ ({\sl dot dashed}), and $\gamma=0.4$ ({\sl thick}).}\label{holographicJJ}
\end{center}
\end{figure}

 To evaluate equation (\ref{phase2}) while eliminating divergences, we follow the methodology outlined by Takayanagi and Fujita in their seminal works \cite{Takayanagi:2011zk, Fujita:2011fp}. By applying these techniques, we can effectively solve equation (\ref{currentjsp}) numerically, as presented in Fig. \ref{holographicJJ}, ensuring that the results are accurate and free from singularities.

In Fig. \ref{p1}, the constricted Josephson junction exhibits microscopic behavior, which is similar to the case of \cite{A.Baselmans}, where the weak electron component ($\Sigma$) responsible for the supercurrent ($J/J_{\max}$) facilitates the transport of correlated electrons. This junction shows Superconductor/Normal/Superconductor (SNS) behavior, with conduction electrons mediating transport through a critical current \cite{Chandrasekhar:2024epu}. Experimentally, as noted by \cite{A.Baselmans}, applying a voltage alters the distribution function within the conductor wire, thereby changing the occupation of the energy levels that carry the supercurrent. In addition, the supercurrent decreases with temperature, defining the critical current at which superconducting materials retain their properties. At higher temperatures, superconductors transition to a normal state, suppressing the manifestation of the Josephson current and eliminating any detectable supercurrent \cite{Chandrasekhar:2024epu}.
The real Josephson junction devices have confirmed that the contribution of mass to the observed effects is negligible \cite{Senapati}. This observation is related to the Horndeski parameters, where the breaking angle $\theta=\theta'=2\pi/3$ results in a thinning of the planar Josephson junction. Energy dispersive X-ray (EDX) elemental maps \cite{Senapati} reveal that several planar Josephson junctions of the SQUID type were fabricated \cite{Jeon,Bhatia,Robinson2007,Robinson2006} following the EDAX protocol. The temperature-dependent resistance $R(\Omega)$ of these junctions is depicted in the right panel of Fig. \ref{holographicJJ}, which aligns with experimental expectations \cite{Senapati}. In particular, precise changes in junction resistance occur with increased grinding time from 6.8 to 9.52 s. Consequently, the supercurrent decreases as the resistance of the junction increases with longer grinding times, corresponding to greater grinding depths.

The geometric coupling illustrated in Fig. \ref{p1} provides insight into the constricted Josephson junction, a planar junction that has revealed intricate phenomena \cite{Senapati,Jeon,Bhatia,Robinson2007,Robinson2006}. The (Cu/Pt)-Nb exemplifies this structure \cite{Bhatia,Robinson2007,Robinson2006}, where our holographic model implies that niobium superconductors simulate the copper/platinum barrier using the boundary tension $\Sigma$. The coupling of the condensate or the interaction between superconducting Cooper pair condensates significantly affects the material properties of the barrier ($\Sigma$), thereby affecting the critical current and the overall behavior of the junction. To show the behavior of the condensate, we extract the expectation value of the operator $\mathcal{O}$, which is given by
\begin{eqnarray}
\frac{<\mathcal{O}>_{y=0,J=0}}{T^{2}_{c}}=\frac{A_{1}}{\sqrt{|-\xi|}}e^{-\frac{\Sigma\,u_h}{2\zeta}},\label{operator1}
\end{eqnarray}
where $\xi={2(\alpha+\gamma\Lambda)}/{\alpha^2}$ is a real parameter, 
 defined in Eq. \eqref{xi}. 

\begin{figure}[!ht]
\begin{center}
\includegraphics[scale=0.55]{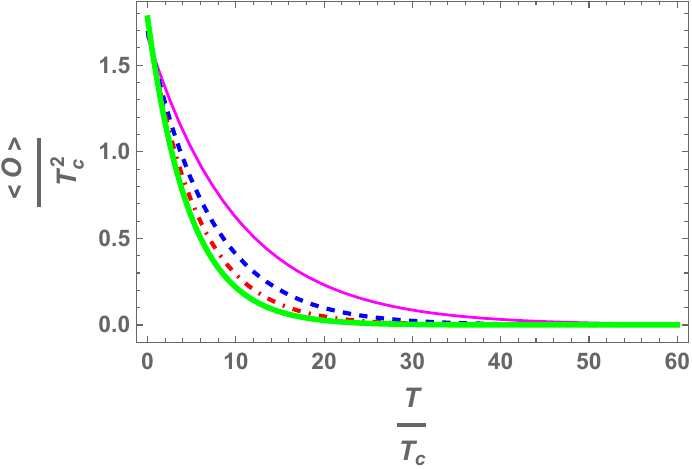}
\includegraphics[scale=0.55]{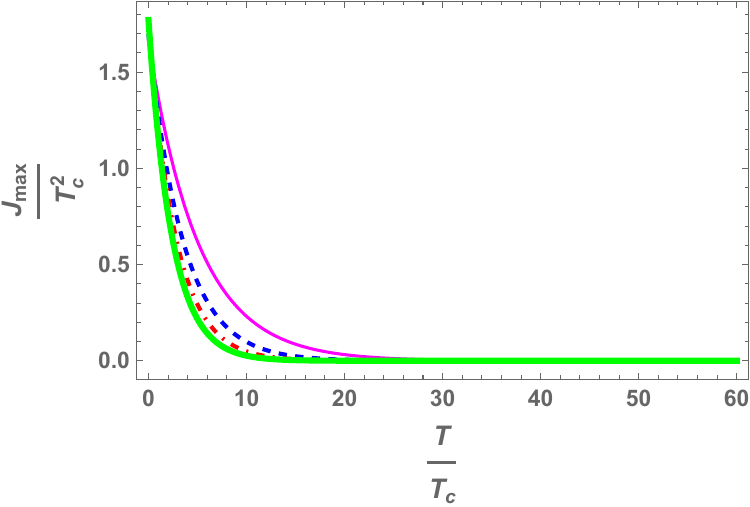}
\caption{The figure shows the behavior of the equations (\ref{operator1}) and (\ref{operator2}) for the values $A_{0}=A_{1}=1$, and $\Sigma=1/4$, $\Lambda=-1$, $\alpha=8/3$ with $\gamma=-0.1$ ({\sl  solid pink}), $\gamma=-0.2$ ({\sl dot dashed blue}), $\gamma=-0.3$ ({\sl dashed red})  and $\gamma=-0.4$ ({\sl thick green}), respectively.}\label{magnetichgxx}
\end{center}
\end{figure}

Note that $\zeta$ is the coherence length \cite{Horowitz:2011dz,Santos:2024cwf}, which is defined here as
\begin{eqnarray}
\zeta\equiv\sqrt{\frac{(\alpha+\gamma\Lambda)}{6\alpha\gamma}}.
\end{eqnarray}
This shows the dependence of the Josephson current $J$ on the Horndeski parameters; see Fig. \ref{holographicJJ}. This coherence length measures the distance over which the superconducting order parameter (which describes the superconducting state) remains correlated. So, if $\gamma$ is large or $\alpha$ is small, the coherence is small (Fig. \ref{magnetichgxx}), highlighting the fact that $\zeta<<\Sigma$, and this fact guaranties the universal law (\ref{operator1}). The curve aligns well with the experimental observations \cite{Hovhannisyan:2022gkp}, which predict exponential behavior for Josephson junctions: 
\begin{eqnarray}
\frac{J_{max}}{T^{2}_{c}}=\frac{A_{0}}{\sqrt{|-\xi|}}e^{\frac{-\Sigma u_{h}}{\zeta}}.\label{operator2}
\end{eqnarray}
This agreement underscores the validity of our model in capturing the physical properties of these systems.

The behavior presented in Fig. \ref{magnetichgxx}, based on equations (\ref{operator1})-(\ref{operator2}) with respect to the parameters $\gamma$ and $\Sigma$, respectively, aligns with both theoretical studies on the SNS junctions \cite{Santos:2024cwf,Siano,Chandrasekhar:2024epu,Senapati,Erdmenger:2014xya,Minamitsuji:2013vra,Kawamoto:2023wzj,Kanda:2023zse,Horowitz:2011dz} and experimental findings \cite{Hovhannisyan:2022gkp,Santos:2024cwf,Senapati,Jeon,Bhatia,Robinson2007,Robinson2006}. The relative magnitude of the supercurrent in the voltage state range is smaller than expected, as noted in \cite{Santos:2024cwf,Horowitz:2011dz}. This discrepancy is attributed to the parameters $\alpha$ and $\gamma$, which modulate the strength of the scalar field in the constricting junction, as geometrically represented in Fig. \ref{p1}. In our model, the junction width is set to $\Sigma=1/4$. Consequently, the distribution function is not uniform throughout the length of the junction, leading to a reduced magnitude of supercurrent \cite{Sickinger}, consistent with theoretical predictions \cite{A.Baselmans}.

\section{Profile description of the normal Josephson junctions}\label{prof2}

\begin{figure}[!ht]
\begin{center}
\includegraphics[width=\textwidth]{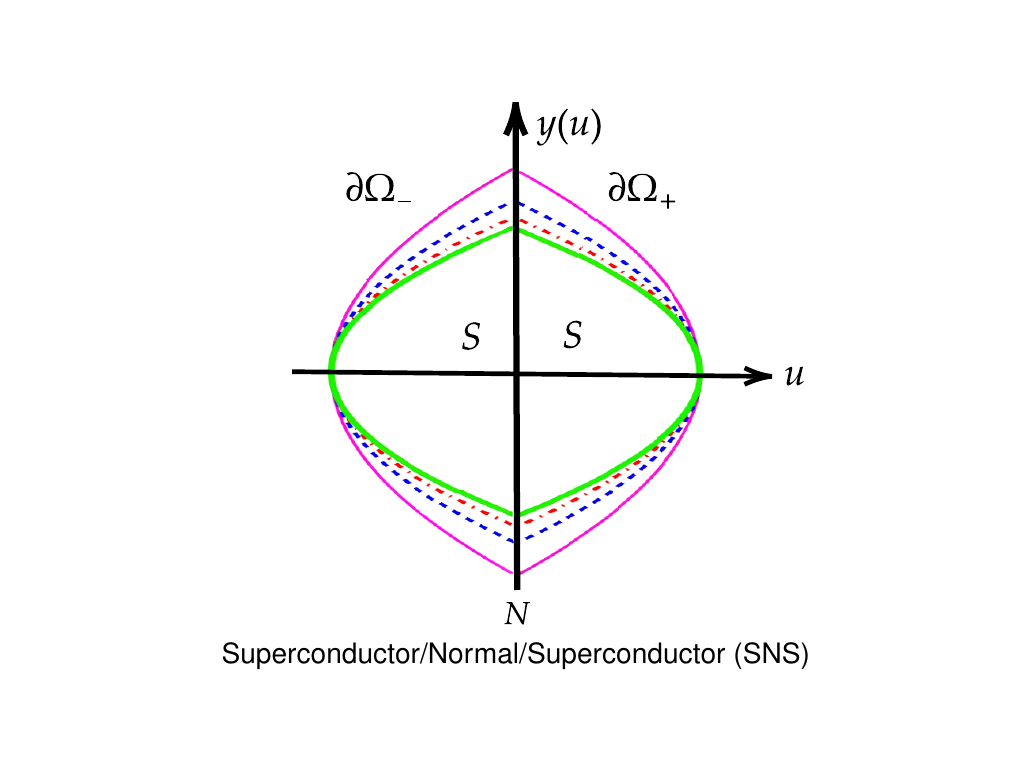}
\caption{The figure shows the behavior of the equation (\ref{scalar5}) for the values $\Lambda=-1$, $\alpha=8/3$ with $\gamma=-0.1$ ({\sl  solid pink}), $\gamma=-0.2$ ({\sl dot dashed blue}), $\gamma=-0.3$ ({\sl dashed red})  and $\gamma=-0.4$ ({\sl thick green}), respectively.}\label{norjunc}
\end{center}
\end{figure}

In this section, we detail  the construction of the normal Josephson junctions in Fig. \ref{p1}, joining the scalar fields $\phi=\phi_{-}\cup\,\phi_{+}$ at $\mathcal{M}_{\pm}=\mathcal{P}_{\pm}$, \cite{Erdmenger:2014xya,Minamitsuji:2013vra,Kawamoto:2023wzj,Kanda:2023zse,Horowitz:2011dz}. The junction condition is:
\begin{eqnarray}
\left[\left(1+\frac{\gamma}{2}(R_h+K^{\rho\sigma}K_{\rho\sigma}-K^2)\right)\phi{'}+\kappa\,S^{\partial\Omega}_{\mu\nu}\,\phi_{|\,\mu\nu}\right]_{\phi_{-}\cup\,\phi_{+}}=\mathcal{F}^{\partial\Omega}_{\phi}\label{scalar1}
\end{eqnarray}
In this analysis, we observe that the covariant derivatives of the scalar field relative to the intrinsic metric $h_{\mu\nu}$ are denoted as $\phi_{|\,\mu\nu}$. In addition, intrinsic curvature terms, such as $R_h$, remain continuous across the surface where $\phi=\phi_{-}\cup\,\phi_{+}$. To avoid the no-hair theorem \cite{Rinaldi:2012vy,Babichev:2013cya,Anabalon:2013oea}, we impose the condition ${\cal E}_{\phi}[g_{rr},\phi]={\cal E}_{rr}[g_{rr},\phi]=0$, which ensures that ${\cal F}^{\partial\Omega}_{\phi}=0$. This leads us to the following.
\begin{eqnarray}
\left[\left(1+\frac{\gamma}{2}(R_h+K^{\rho\sigma}K_{\rho\sigma}-K^2)\right)\phi{'}+\kappa\,S^{\partial\Omega}_{\mu\nu}\,\phi_{|\,\mu\nu}\right]_{\phi_{-}\cup\,\phi_{+}}=0.\label{scalar2}
\end{eqnarray}
Furthermore, let us note that $S^{\partial\Omega}_{\mu\nu}$ is given by the equation \eqref{H7} and, for consistency, we again have $S^{\partial\Omega}_{\mu\nu}=0$ and thus we are left with the equation that describes the scalar field responsible for the junction profile, which simplifies to:
\begin{eqnarray}
\left[\left(1+\frac{\gamma}{2}(R_h+K^{\rho\sigma}K_{\rho\sigma}-K^2)\right)\phi{'}\right]_{\phi_{-}\cup\,\phi_{+}}=0.\label{scalar3}
\end{eqnarray}
This condition ensures that the scalar-field configuration remains consistent with the boundary requirements, allowing us to focus on the intrinsic dynamics of the junction. Through the non-zero components of the Einstein tensor \cite{Santos:2025fdp,Minamitsuji:2013vra}:
\begin{eqnarray}
&&G_{yy}=-\frac{1}{2}R_h-\frac{1}{2}K^{\alpha\beta}K_{\alpha\beta}+\frac{1}{2}K^2,\nonumber\\
&&G_{\alpha\beta}=-K_{\alpha\beta,y}+G^{(h)}_{\alpha\beta}+2K_{\alpha\gamma}K^{\gamma}_{\,\beta}+\frac{1}{2}h_{\alpha\beta}(K^2+K^{\rho\sigma}K_{\rho\sigma})+h_{\alpha\beta}K_{,y}\,,\nonumber\\
&&G_{y\mu}=K^{\alpha}_{\mu|\alpha}-K_{|\mu},
\end{eqnarray}
we can rewrite equation (\ref{scalar3}) as
\begin{eqnarray}
(1-\gamma\,G_{yy})\phi{'}|_{y=0^{\pm}}=0,\label{scalar4}
\end{eqnarray}
with boundary conditions at $u\to\infty$, we have the solution
\begin{eqnarray}
y(u)=\frac{3}{4\alpha}\sqrt{\frac{\gamma}{\alpha}}\left(1+u+\frac{1}{u}\right).\label{scalar5}
\end{eqnarray}
With this profile, we construct a Superconductor-Normal-Superconductor (SNS) junction, as illustrated in Fig. \ref{norjunc}. In this configuration, we achieve highly transparent superconductor-normal interfaces, improving the well-known Andreev reflection process \cite{Haxell:2023uci, Paudel:2024xyy, Hinderling:2024jfq}. Within this medium, electrons are coherently retroreflected as holes with reversed spin and momentum while simultaneously transferring a Cooper pair to the superconductor \cite{Hinderling:2024jfq}. This process is crucial for understanding the transport properties and coherence effects of SNS junctions.

\begin{figure}[!ht]
\begin{center}
\includegraphics[scale=0.55]{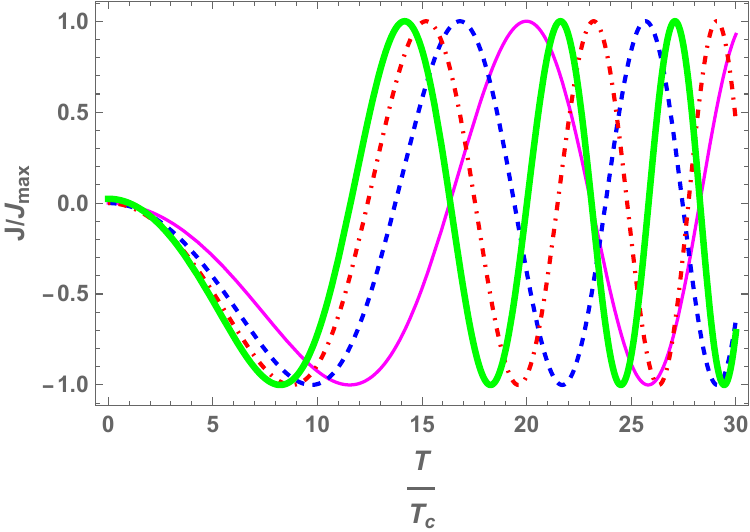}
\includegraphics[scale=0.55]{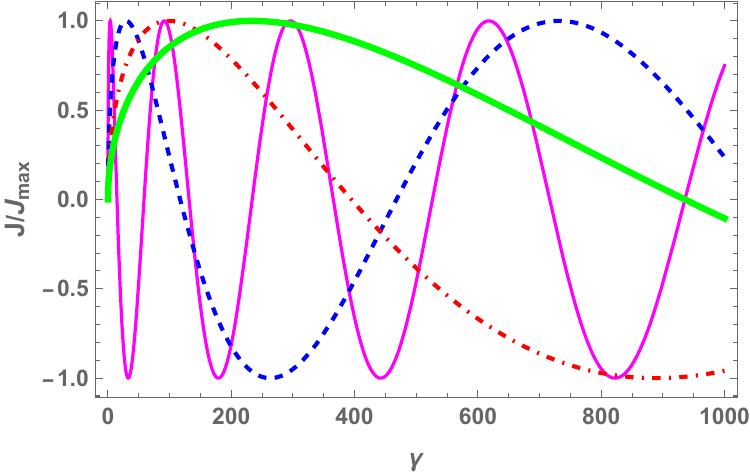}
\caption{This figure shows the behavior of the equation (\ref{currentjsp}). In the left side for the values, $\Lambda=-1$, $\alpha=-8/3$, with $\gamma=-0.1$ ({\sl solid}), $\gamma=-0.2$ ({\sl dashed}), $\gamma=-0.3$ ({\sl dot dashed}), and $\gamma=-0.4$ ({\sl thick}). In the right side is for the values of $T/T_c=0.4$ with $\alpha=0.1$ ({\sl solid}), $\alpha=0.2$ ({\sl dashed}), $\alpha=0.3$ ({\sl dot dashed}), and $\alpha=0.4$ ({\sl thick}).}\label{holographicJJ1}
\end{center}
\end{figure}

To evaluate the Josephson current behavior for the SNS junction, we compute the phase difference:
\begin{eqnarray}
\Gamma&=&-\int^{+\infty}_{-\infty}{du\frac{y^{'}(u)}{u}}\nonumber\\
      &=&-\int^{-u_h}_{-\infty}{\frac{3}{4\alpha}\sqrt{\frac{\gamma}{\alpha}}\left(1+u+\frac{1}{u}\right)du}+\int^{\infty}_{u_h}{\frac{3}{4\alpha}\sqrt{\frac{\gamma}{\alpha}}\left(1+u+\frac{1}{u}\right)du}.\label{phase3}
\end{eqnarray}
By solving this equation, we have the following:
\begin{eqnarray}
\Gamma=\frac{1}{4}\frac{3}{4\alpha}\sqrt{\frac{\gamma}{\alpha}}\,u^{2}_{h}.\label{phase4}
\end{eqnarray}
We observe that while the holographic method applies to any Josephson junction \cite{Santos:2024cwf,Horowitz:2011dz,Hovhannisyan:2022gkp}, the Horndeski scalar field, controlled by parameters $\alpha$ and $\gamma$, must be sufficiently small to enhance the sensitivity of the junction \cite{Hovhannisyan:2022gkp}. This sensitivity improves the observation of the tunneling current, as shown in Fig. \ref{holographicJJ1}. For fixed $\alpha$, the sensitivity of the Josephson current to variations in the $\gamma$ parameter remains significant at low temperatures. The tunneling current remains observable as we increase the force of the fixed $\gamma$ parameter in Fig. \ref{holographicJJ1}. We can see that even at low temperatures, the fact that $\gamma$ is very large decreases the coherence length, allows the current to persist, and reduces thermal noise. However, other sources of decoherence, such as electromagnetic noise \cite{Marcos:2013aya}, material defects, and quasiparticle excitations, can still affect the system \cite{Simmonds:2004ucc}.

Through Horndeski gravity, we demonstrate the existence of analytical solutions for the condensate and current values using the AdS/BCFT correspondence. These solutions provide valuable insights into the behavior of the condensate in this framework, highlighting the potential of Horndeski gravity to model complex interactions in holographic systems. Explicitly, 
\begin{eqnarray}
\frac{<\mathcal{O}>_{y=0,J=0}}{T^{2}_{c}}&=&\frac{A_{1}}{\sqrt{|-\xi|}}e^{-\frac{3}{4\alpha}\sqrt{\frac{\gamma}{\alpha}}u^2_h},\label{OP1}\\
\frac{J_{max}}{T^{2}_{c}}&=&\frac{A_{0}}{\sqrt{|-\xi|}}e^{-\frac{3}{4\alpha}\sqrt{\frac{\gamma}{\alpha}}u^2_h}\,,
\label{OP2}
\end{eqnarray}
where $\xi={2(\alpha+\gamma\Lambda)}/{\alpha^2}$, and the coherence length in this case is
\begin{eqnarray}
\zeta &\equiv& \frac{1}{\frac{3}{4\alpha}\sqrt{\frac{\gamma}{\alpha}}}.
\end{eqnarray}

\begin{figure}[!ht]
\begin{center}
\includegraphics[scale=0.55]{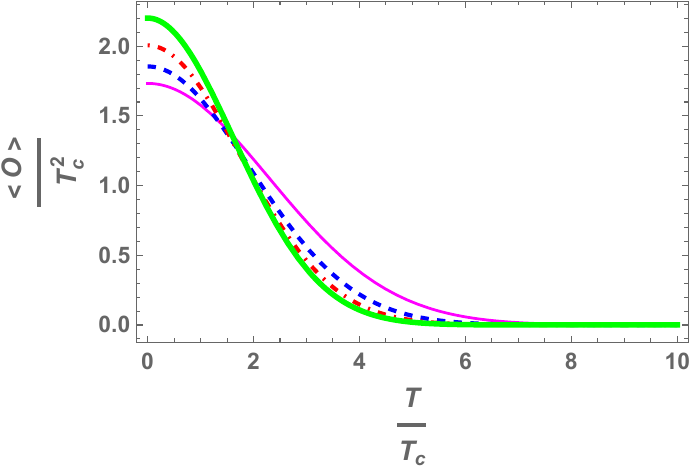}
\includegraphics[scale=0.55]{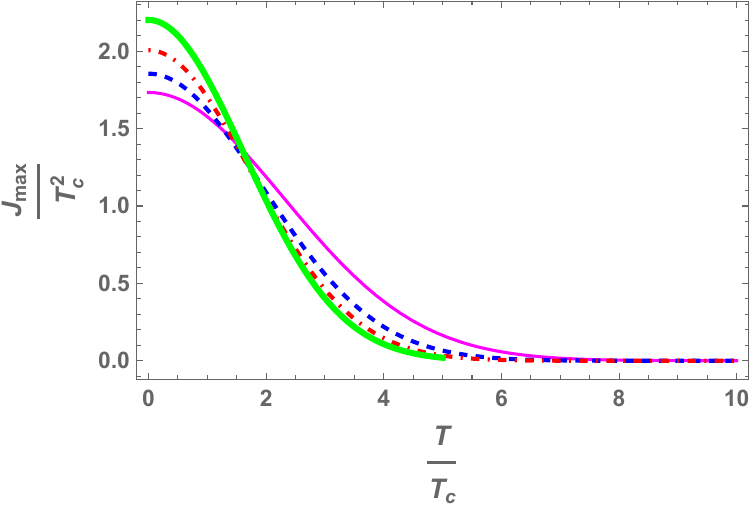}
\caption{The figure shows the behavior of equations (\ref{OP1}) and (\ref{OP2}) for the values $A_{0}=A_{1}=1$, $\Lambda=-1$, $\alpha=8/3$ with $\gamma=-0.1$ ({\sl solid}), $\gamma=-0.2$ ({\sl dashed}), $\gamma=-0.3$ ({\sl dot dashed}), and $\gamma=-0.4$ ({\sl thick}).}\label{conden1}
\end{center}
\end{figure}

Horndeski gravity offers a dual holographic description of superconductors. The curves shown in Fig. \ref{conden1} qualitatively resemble those observed in various materials \cite{Hartnoll:2008vx,Senapati,A.Baselmans,Jeon,Bhatia,Robinson2007,Robinson2006,Sickinger}, in which the condensate approaches a constant value at zero temperature. This behavior underscores Horndeski gravity in modeling the superconducting phase transition and provides insight into the universal properties of superconductors. The advantage of using Horndeski gravity in the construction of SNS junctions via the AdS/BCFT correspondence is that, at low temperatures, the condensate remains finite. This behavior ensures that the system stays within the validity region of our approximation, providing a robust framework for analyzing superconducting properties without encountering divergences.

\section{Conclusions and discussions}\label{concl}

In this work, we show that the AdS$_4$/BCFT$_3$ correspondence within Horndeski gravity allows a simple (3+1)-dimensional bulk theory to replicate several properties of a (2+1)-dimensional superconductor, applicable to both constricted and SNS-normal Josephson junctions. In particular, this constitutes, to our knowledge, the first explicit realization of both constriction‑type and SNS‑type Josephson junctions in an AdS/BCFT setup with Horndeski couplings, thereby extending previous AdS/BCFT–Horndeski applications to genuine superconducting transport phenomena. A key advantage of using Horndeski gravity is that it provides appropriate gravitational solutions via the no-hair theorem, which improves the Josephson effect on both the boundary side of the constriction junction and the gravitational side of the normal junction.

We provide a direct writing method for constructing geometric Josephson constriction junctions \cite{Santos:2024cwf} within the framework of Horndeski gravity, where the spatial dependence of the profile $y(u)$ does not lead to the vanishing of $J$. The superfluid current is proportional to the sine of the phase difference, which is controlled by the Horndeski parameters at the boundary $\partial\Omega$. The constriction junction is characterized by an exponential decay in supercurrent with increasing gap width $\Sigma$ and parameter $\gamma$ at $J_{max}$. Our findings suggest that the DC Josephson holographic junction model is applicable to the case of the AdS$_4$/BCFT$_3$ correspondence with Horndeski gravity. The microscopic mechanism responsible for the supercurrent in a Josephson junction with constriction involves the $\Sigma$ tension at the boundary $\partial\Omega$. The gauge fields introduce a voltage difference $V$, effectively adding a constant $V$ to $A_{t}$, corresponding to a phase shift in the charged scalar $\Psi$ by $qVt$. The geometric Josephson junctions presented here within Horndeski gravity ensure that the superconducting state parameters adjust the junction's transport measurements. Although this division may seem somewhat artificial, a clear distinction is often evident based on voltage characteristics \cite{Ruhtinas}.

In the fabrication of Josephson SNS junctions, the weak link within the junction is highly tunable by varying the Horndeski parameters. These parameters characterize the "metallic" nature of the link as a result of the superconducting transition. We have successfully proposed SNS devices that achieve critical current densities and junction resistances. The microscopic mechanism responsible for the supercurrent in this type of Josephson junction is the transport of correlated electrons. In Fig. \ref{holographicJJ1}, conduction electrons in an SNS junction mediate current transport from superconductor 1 ($S_1$) to superconductor 2 ($S_2$), similar to ballistic or diffusive transport through the normal metal ($N$) \cite{A.Baselmans}.

\acknowledgments

Fabiano F. Santos is partially supported by Conselho Nacional de Desenvolvimento Cient\'{\i}fico e Tecnol\'{o}gico (CNPq) under grant 302835/2024-5. 
This work was supported by Coordenação de Aperfeiçoamento de Pessoal de Nível Superior (CAPES) under finance code 001. 
HBF is partially supported by Conselho Nacional de Desenvolvimento Cient\'{\i}fico e Tecnol\'{o}gico (CNPq) under grant  310346/2023-1, and Fundação Carlos Chagas Filho de Amparo à Pesquisa do Estado do Rio de Janeiro (FAPERJ) under grant E-26/204.095/2024.



\begin{thebibliography}{99}



\bibitem{Josephson:1962zz}
B.~D.~Josephson,
{\it Possible new effects in superconductive tunnelling},
Phys. Lett. \textbf{1}, 251-253 (1962)
doi:10.1016/0031-9163(62)91369-0

\bibitem{Hovhannisyan:2022gkp}
R.~A.~Hovhannisyan, T.~Golod and V.~M.~Krasnov,
{\it Holographic reconstruction of magnetic field distribution in a Josephson junction from diffraction-like Ic(H) patterns},''
Phys. Rev. B \textbf{105}, no.21, 214513 (2022)
doi:10.1103/PhysRevB.105.214513
[arXiv:2205.12837 [cond-mat.supr-con]].






\bibitem{Seiberg:1994rs}
N.~Seiberg and E.~Witten,
``Electric - magnetic duality, monopole condensation, and confinement in N=2 supersymmetric Yang-Mills theory,''
Nucl. Phys. B \textbf{426}, 19-52 (1994)
[erratum: Nucl. Phys. B \textbf{430}, 485-486 (1994)]
doi:10.1016/0550-3213(94)90124-4
[arXiv:hep-th/9407087 [hep-th]].

\bibitem{Witten:1995ex}
E.~Witten,
``String theory dynamics in various dimensions,''
Nucl. Phys. B \textbf{443}, 85-126 (1995)
doi:10.1016/0550-3213(95)00158-O
[arXiv:hep-th/9503124 [hep-th]].


  \bibitem{Maldacena:1997re} 
  J.~M.~Maldacena,
  {\it The Large N limit of superconformal field theories and supergravity},
  Int.\ J.\ Theor.\ Phys.\  {\bf 38}, 1113 (1999)
  [Adv.\ Theor.\ Math.\ Phys.\  {\bf 2}, 231 (1998)]
    [hep-th/9711200].

\bibitem{Gubser:1998bc}
S.~S.~Gubser, I.~R.~Klebanov and A.~M.~Polyakov,
``Gauge theory correlators from noncritical string theory,''
Phys. Lett. B \textbf{428}, 105-114 (1998)
doi:10.1016/S0370-2693(98)00377-3
[arXiv:hep-th/9802109 [hep-th]].
  
\bibitem{Witten:1998qj}
E.~Witten,
{\it Anti-de Sitter space and holography},
Adv. Theor. Math. Phys. \textbf{2} (1998), 253-291
doi:10.4310/ATMP.1998.v2.n2.a2
[arXiv:hep-th/9802150 [hep-th]].

\bibitem{Aharony:1999ti}
O.~Aharony, S.~S.~Gubser, J.~M.~Maldacena, H.~Ooguri and Y.~Oz,
``Large N field theories, string theory and gravity,''
Phys. Rept. \textbf{323}, 183-386 (2000)
doi:10.1016/S0370-1573(99)00083-6
[arXiv:hep-th/9905111 [hep-th]].



\bibitem{Son:2007vk}
D.~T.~Son and A.~O.~Starinets,
``Viscosity, Black Holes, and Quantum Field Theory,''
Ann. Rev. Nucl. Part. Sci. \textbf{57}, 95-118 (2007)
doi:10.1146/annurev.nucl.57.090506.123120
[arXiv:0704.0240 [hep-th]].


\bibitem{Erdmenger:2007cm}
J.~Erdmenger, N.~Evans, I.~Kirsch and E.~Threlfall,
``Mesons in Gauge/Gravity Duals - A Review,''
Eur. Phys. J. A \textbf{35}, 81-133 (2008)
doi:10.1140/epja/i2007-10540-1
[arXiv:0711.4467 [hep-th]].

\bibitem{Schafer:2009dj}
T.~Sch{\"a}fer and D.~Teaney,
``Nearly Perfect Fluidity: From Cold Atomic Gases to Hot Quark Gluon Plasmas,''
Rept. Prog. Phys. \textbf{72}, 126001 (2009)
doi:10.1088/0034-4885/72/12/126001
[arXiv:0904.3107 [hep-ph]].

\bibitem{Rangamani:2009xk}
M.~Rangamani,
``Gravity and Hydrodynamics: Lectures on the fluid-gravity correspondence,''
Class. Quant. Grav. \textbf{26}, 224003 (2009)
doi:10.1088/0264-9381/26/22/224003
[arXiv:0905.4352 [hep-th]].


\bibitem{Casalderrey-Solana:2011dxg}
J.~Casalderrey-Solana, H.~Liu, D.~Mateos, K.~Rajagopal and U.~Achim Wiedemann,
``Gauge/String Duality, Hot QCD and Heavy Ion Collisions,''
Cambridge University Press, 2014,
ISBN 978-1-009-40350-4, 978-1-009-40349-8, 978-1-009-40352-8, 978-1-139-13674-7
doi:10.1017/9781009403504
[arXiv:1101.0618 [hep-th]].


\bibitem{Hartnoll:2016apf}
S.~A.~Hartnoll, A.~Lucas and S.~Sachdev,
``Holographic quantum matter,''
[arXiv:1612.07324 [hep-th]].

\bibitem{Florkowski:2017olj}
W.~Florkowski, M.~P.~Heller and M.~Spalinski,
``New theories of relativistic hydrodynamics in the LHC era,''
Rept. Prog. Phys. \textbf{81}, no.4, 046001 (2018)
doi:10.1088/1361-6633/aaa091
[arXiv:1707.02282 [hep-ph]].

\bibitem{Hartnoll:2008vx}
S.~A.~Hartnoll, C.~P.~Herzog and G.~T.~Horowitz,
{\it Building a Holographic Superconductor},
Phys. Rev. Lett. \textbf{101}, 031601 (2008)
doi:10.1103/PhysRevLett.101.031601
[arXiv:0803.3295 [hep-th]].

\bibitem{Arean:2010xd}
D.~Arean, M.~Bertolini, J.~Evslin and T.~Prochazka,
{\it On Holographic Superconductors with DC Current},
JHEP \textbf{07}, 060 (2010)
doi:10.1007/JHEP07(2010)060
[arXiv:1003.5661 [hep-th]].

\bibitem{Horowitz:2011dz}
G.~T.~Horowitz, J.~E.~Santos and B.~Way,
{\it A Holographic Josephson Junction},
Phys. Rev. Lett. \textbf{106}, 221601 (2011)
doi:10.1103/PhysRevLett.106.221601
[arXiv:1101.3326 [hep-th]].

\bibitem{Wang:2011rva}
Y.~Q.~Wang, Y.~X.~Liu and Z.~H.~Zhao,
{\it Holographic Josephson Junction in 3+1 dimensions},
[arXiv:1104.4303 [hep-th]].

\bibitem{Wang:2012yj}
Y.~Q.~Wang, Y.~X.~Liu, R.~G.~Cai, S.~Takeuchi and H.~Q.~Zhang,
{\it Holographic SIS Josephson Junction},
JHEP \textbf{09}, 058 (2012)
doi:10.1007/JHEP09(2012)058
[arXiv:1205.4406 [hep-th]].

\bibitem{Domokos:2012rj}
S.~K.~Domokos, C.~Hoyos and J.~Sonnenschein,
{\it Holographic Josephson Junctions and Berry holonomy from D-branes},
JHEP \textbf{10}, 073 (2012)
doi:10.1007/JHEP10(2012)073
[arXiv:1207.2182 [hep-th]].

\bibitem{Hu:2015dnl}
Y.~P.~Hu, H.~F.~Li, H.~B.~Zeng and H.~Q.~Zhang,
{\it Holographic Josephson Junction from Massive Gravity},
Phys. Rev. D \textbf{93}, no.10, 104009 (2016)
doi:10.1103/PhysRevD.93.104009
[arXiv:1512.07035 [hep-th]].



\bibitem{Takayanagi:2011zk}{ 
T.~Takayanagi,
{\it Holographic Dual of BCFT},
Phys.\ Rev.\ Lett.\  {\bf 107}, 101602 (2011),
[arXiv:1105.5165 [hep-th]].}
doi:10.1103/PhysRevLett.107.101602;


\bibitem{Fujita:2011fp}{ 
M.~Fujita, T.~Takayanagi and E.~Tonni,
{\it Aspects of AdS/BCFT},
JHEP {\bf 1111}, 043 (2011),
[arXiv:1108.5152 [hep-th]].}
doi:10.1007/JHEP11(2011)043;

\bibitem{Fujita:2012fp}
M.~Fujita, M.~Kaminski and A.~Karch,
{\it SL(2,Z) Action on AdS/BCFT and Hall Conductivities},
JHEP \textbf{07}, 150 (2012)
doi:10.1007/JHEP07(2012)150
[arXiv:1204.0012 [hep-th]].

	
\bibitem{Melnikov:2012tb}{ 
  D.~Melnikov, E.~Orazi and P.~Sodano,
  {\it On the AdS/BCFT Approach to Quantum Hall Systems},
  JHEP {\bf 1305}, 116 (2013),
  [arXiv:1211.1416 [hep-th]].}
doi:10.1007/JHEP05(2013)116;

  \bibitem{Magan:2014dwa}{ 
  J.~M.~Magán, D.~Melnikov and M.~R.~O.~Silva,
  {\it Black Holes in AdS/BCFT and Fluid/Gravity Correspondence},
  JHEP {\bf 1411}, 069 (2014),
  [arXiv:1408.2580 [hep-th]].}
doi:10.1007/JHEP11(2014)069;

    
\bibitem{Cavalcanti:2018pta} 
  A.~G.~Cavalcanti, D.~Melnikov and M.~R.~O.~Silva,
  {\it Studies of Boundary Entropy in AdS/BCFT},
  arXiv:1808.07966 [hep-th].
	

\bibitem{Santos:2023flb}
F.~F.~Santos, M.~Bravo-Gaete, O.~Sokoliuk and A.~Baransky,
``AdS/BCFT Correspondence and Horndeski Gravity in the Presence of Gauge Fields: Holographic Paramagnetism/Ferromagnetism Phase Transition,''
Fortsch. Phys. \textbf{71}, no.12, 2300008 (2023)
doi:10.1002/prop.202300008
[arXiv:2301.03121 [hep-th]].


\bibitem{Erdmenger:2015spo}
J.~Erdmenger, M.~Flory, C.~Hoyos, M.~N.~Newrzella and J.~M.~S.~Wu,
{\it Entanglement Entropy in a Holographic Kondo Model},
Fortsch. Phys. \textbf{64}, 109-130 (2016)
doi:10.1002/prop.201500099
[arXiv:1511.03666 [hep-th]].

\bibitem{Bachas:2020yxv}
C.~Bachas, S.~Chapman, D.~Ge and G.~Policastro,
{\it Energy Reflection and Transmission at 2D Holographic Interfaces},
Phys. Rev. Lett. \textbf{125}, no.23, 231602 (2020)
doi:10.1103/PhysRevLett.125.231602
[arXiv:2006.11333 [hep-th]].

\bibitem{Bachas:2021fqo}
C.~Bachas and V.~Papadopoulos,
{\it Phases of Holographic Interfaces},
JHEP \textbf{04}, 262 (2021)
doi:10.1007/JHEP04(2021)262
[arXiv:2101.12529 [hep-th]].

\bibitem{Aharony:2003qf}
O.~Aharony, O.~DeWolfe, D.~Z.~Freedman and A.~Karch,
{\it Defect conformal field theory and locally localized gravity},
JHEP \textbf{07}, 030 (2003)
doi:10.1088/1126-6708/2003/07/030
[arXiv:hep-th/0303249 [hep-th]].

\bibitem{Bak:2003jk}
D.~Bak, M.~Gutperle and S.~Hirano,
{\it A Dilatonic deformation of AdS(5) and its field theory dual},
JHEP \textbf{05}, 072 (2003)
doi:10.1088/1126-6708/2003/05/072
[arXiv:hep-th/0304129 [hep-th]].

\bibitem{DHoker:2007zhm}
E.~D'Hoker, J.~Estes and M.~Gutperle,
{\it Exact half-BPS Type IIB interface solutions. I. Local solution and supersymmetric Janus},
JHEP \textbf{06}, 021 (2007)
doi:10.1088/1126-6708/2007/06/021
[arXiv:0705.0022 [hep-th]].

\bibitem{Bak:2007jm}
D.~Bak, M.~Gutperle and S.~Hirano,
{\it Three dimensional Janus and time-dependent black holes},
JHEP \textbf{02}, 068 (2007)
doi:10.1088/1126-6708/2007/02/068
[arXiv:hep-th/0701108 [hep-th]].

\bibitem{Azeyanagi:2007qj}
T.~Azeyanagi, A.~Karch, T.~Takayanagi and E.~G.~Thompson,
{\it Holographic calculation of boundary entropy},
JHEP \textbf{03}, 054 (2008)
doi:10.1088/1126-6708/2008/03/054
[arXiv:0712.1850 [hep-th]].


\bibitem{Liu:2024oxg}
Y.~Liu, H.~D.~Lyu and C.~Y.~Wang,
{\it On AdS$_{3}$/ICFT$_{2}$ with a dynamical scalar field located on the brane},
JHEP \textbf{10}, 001 (2024)
doi:10.1007/JHEP10(2024)001
[arXiv:2403.20102 [hep-th]].

\bibitem{Liu:2022ezb}
Y.~Liu, H.~D.~Lyu and J.~K.~Zhao,
{\it Properties of gapped systems in AdS/BCFT},
Phys. Rev. D \textbf{107}, no.6, 066017 (2023)
doi:10.1103/PhysRevD.107.066017
[arXiv:2210.02802 [hep-th]].





\bibitem{Chen:2020uac}
H.~Z.~Chen, R.~C.~Myers, D.~Neuenfeld, I.~A.~Reyes and J.~Sandor,
{\it Quantum Extremal Islands Made Easy, Part I: Entanglement on the Brane},
JHEP \textbf{10}, 166 (2020)
doi:10.1007/JHEP10(2020)166
[arXiv:2006.04851 [hep-th]].

\bibitem{Chen:2020hmv}
H.~Z.~Chen, R.~C.~Myers, D.~Neuenfeld, I.~A.~Reyes and J.~Sandor,
{\it Quantum Extremal Islands Made Easy, Part II: Black Holes on the Brane},
JHEP \textbf{12}, 025 (2020)
doi:10.1007/JHEP12(2020)025
[arXiv:2010.00018 [hep-th]].

\bibitem{Geng:2020fxl}
H.~Geng, A.~Karch, C.~Perez-Pardavila, S.~Raju, L.~Randall, M.~Riojas and S.~Shashi,
{\it Information Transfer with a Gravitating Bath},
SciPost Phys. \textbf{10}, no.5, 103 (2021)
doi:10.21468/SciPostPhys.10.5.103
[arXiv:2012.04671 [hep-th]].



\bibitem{Santos:2024cwf}
F.~F.~Santos and H.~Boschi-Filho,
``{\it Geometric Josephson junction},''
JHEP \textbf{01}, 135 (2025)
doi:10.1007/JHEP01(2025)135
[arXiv:2407.10008 [hep-th]].




\bibitem{Horndeski:1974wa}
G.~W.~Horndeski,
``Second-order scalar-tensor field equations in a four-dimensional space,''
Int. J. Theor. Phys. \textbf{10}, 363-384 (1974)
doi:10.1007/BF01807638

\bibitem{Koyama:2015vza}
K.~Koyama,
``Cosmological Tests of Modified Gravity,''
Rept. Prog. Phys. \textbf{79}, no.4, 046902 (2016)
doi:10.1088/0034-4885/79/4/046902
[arXiv:1504.04623 [astro-ph.CO]].

\bibitem{Ishak:2018his}
M.~Ishak,
``Testing General Relativity in Cosmology,''
Living Rev. Rel. \textbf{22}, no.1, 1 (2019)
doi:10.1007/s41114-018-0017-4
[arXiv:1806.10122 [astro-ph.CO]].

\bibitem{Heisenberg:2018vsk}
L.~Heisenberg,
``A systematic approach to generalisations of General Relativity and their cosmological implications,''
Phys. Rept. \textbf{796}, 1-113 (2019)
doi:10.1016/j.physrep.2018.11.006
[arXiv:1807.01725 [gr-qc]].


\bibitem{Kobayashi:2019hrl}
T.~Kobayashi,
``Horndeski theory and beyond: a review,''
Rept. Prog. Phys. \textbf{82}, no.8, 086901 (2019)
doi:10.1088/1361-6633/ab2429
[arXiv:1901.07183 [gr-qc]].




\bibitem{Jiang:2017imk}
W.~J.~Jiang, H.~S.~Liu, H.~Lu and C.~N.~Pope,
{\it DC Conductivities with Momentum Dissipation in Horndeski Theories},
JHEP \textbf{07}, 084 (2017)
doi:10.1007/JHEP07(2017)084
[arXiv:1703.00922 [hep-th]].


\bibitem{Zhang:2022hxl}
D.~Zhang, G.~Fu, X.~J.~Wang, Q.~Pan and J.~P.~Wu,
{\it Transport properties in the Horndeski holographic two-currents model},
Eur. Phys. J. C \textbf{83}, no.4, 316 (2023)
doi:10.1140/epjc/s10052-023-11444-8
[arXiv:2211.07074 [hep-th]].

\bibitem{Lu:2020ttp}
J.~W.~Lu, Y.~B.~Wu, L.~G.~Mi, H.~Liao and B.~P.~Dong,
{\it Holographic s-wave superconductors with Horndeski correction},
Eur. Phys. J. C \textbf{80}, no.7, 605 (2020)
doi:10.1140/epjc/s10052-020-8173-6

\bibitem{Santos:2021orr}
F.~F.~Santos, E.~F.~Capossoli and H.~Boschi-Filho,
{\it AdS/BCFT correspondence and BTZ black hole thermodynamics within Horndeski gravity},
Phys. Rev. D \textbf{104}, no.6, 066014 (2021)
doi:10.1103/PhysRevD.104.066014
[arXiv:2105.03802 [hep-th]].


\bibitem{Santos:2024cvx}
F.~F.~Santos, B.~Pourhassan, E.~N.~Saridakis, O.~Sokoliuk, A.~Baransky and E.~O.~Kahya,
{\it Holographic boundary conformal field theory within Horndeski gravity},
JHEP \textbf{12}, 217 (2025)
doi:10.1007/JHEP12(2024)217
[arXiv:2410.18781 [hep-th]].


\bibitem{Santos:2025wbl}
F.~F.~Santos,
``Probing the Black Hole Interior with Holographic Entanglement Entropy and the Role of AdS/BCFT Correspondence,''
Fortsch. Phys. \textbf{73}, no.9-10, e70031 (2025)
doi:10.1002/prop.70031
[arXiv:2508.21224 [hep-th]].

\bibitem{Santos:2025fdp}
F.~F.~Santos, B.~Pourhassan and E.~N.~Saridakis,
{\it Black Hole Entropy and Complexity Growth in Horndeski Gravity within the AdS/BCFT Framework},
[arXiv:2509.23430 [hep-th]].





\bibitem{Geng:2023iqd}
H.~Geng, A.~Karch, C.~Perez-Pardavila, L.~Randall, M.~Riojas, S.~Shashi and M.~Youssef,
{\it Constraining braneworlds with entanglement entropy},
SciPost Phys. \textbf{15}, no.5, 199 (2023)
doi:10.21468/SciPostPhys.15.5.199
[arXiv:2306.15672 [hep-th]].

\bibitem{Geng:2020qvw}
H.~Geng and A.~Karch,
{\it Massive islands},
JHEP \textbf{09}, 121 (2020)
doi:10.1007/JHEP09(2020)121
[arXiv:2006.02438 [hep-th]].

\bibitem{Geng:2021wcq}
H.~Geng, Y.~Nomura and H.~Y.~Sun,
{\it Information paradox and its resolution in de Sitter holography},
Phys. Rev. D \textbf{103}, no.12, 126004 (2021)
doi:10.1103/PhysRevD.103.126004
[arXiv:2103.07477 [hep-th]].

\bibitem{Geng:2021iyq}
H.~Geng, S.~L{\"u}st, R.~K.~Mishra and D.~Wakeham,
{\it Holographic BCFTs and Communicating Black Holes},
JHEP \textbf{08}, 003 (2021)
doi:10.1007/JHEP08(2021)003
[arXiv:2104.07039 [hep-th]].

\bibitem{Geng:2021mic}
H.~Geng, A.~Karch, C.~Perez-Pardavila, S.~Raju, L.~Randall, M.~Riojas and S.~Shashi,
{\it Entanglement phase structure of a holographic BCFT in a black hole background},
JHEP \textbf{05}, 153 (2022)
doi:10.1007/JHEP05(2022)153
[arXiv:2112.09132 [hep-th]].



\bibitem{Geng:2022dua}
H.~Geng, L.~Randall and E.~Swanson,
{\it BCFT in a black hole background: an analytical holographic model},
JHEP \textbf{12}, 056 (2022)
doi:10.1007/JHEP12(2022)056
[arXiv:2209.02074 [hep-th]].

\bibitem{Geng:2023qwm}
H.~Geng,
{\it Revisiting Recent Progress in the Karch-Randall Braneworld},
[arXiv:2306.15671 [hep-th]].

\bibitem{Geng:2022slq}
H.~Geng, A.~Karch, C.~Perez-Pardavila, S.~Raju, L.~Randall, M.~Riojas and S.~Shashi,
{\it Jackiw-Teitelboim Gravity from the Karch-Randall Braneworld},
Phys. Rev. Lett. \textbf{129}, no.23, 231601 (2022)
doi:10.1103/PhysRevLett.129.231601
[arXiv:2206.04695 [hep-th]].

\bibitem{Geng:2022tfc}
H.~Geng,
{\it Aspects of AdS$_{2}$ quantum gravity and the Karch-Randall braneworld},
JHEP \textbf{09}, 024 (2022)
doi:10.1007/JHEP09(2022)024
[arXiv:2206.11277 [hep-th]].

\bibitem{Geng:2024xpj}
H.~Geng,
{\it Replica wormholes and entanglement islands in the Karch-Randall braneworld},
JHEP \textbf{01}, 063 (2025)
doi:10.1007/JHEP01(2025)063
[arXiv:2405.14872 [hep-th]].

\bibitem{Geng:2025efs}
H.~Geng, L.~Y.~Hung and Y.~Jiang,
{\it from ETH: Multi-interval Entanglement and Replica Wormholes from Large-$c$ BCFT Ensemble},
[arXiv:2505.20385 [hep-th]].

\bibitem{Bao:2025plr}
N.~Bao, H.~Geng and Y.~Jiang,
``Ryu-Takayanagi formula for multi-boundary black holes from 2D large-c CFT ensemble,''
JHEP \textbf{10}, 042 (2025)
doi:10.1007/JHEP10(2025)042
[arXiv:2504.12388 [hep-th]].

\bibitem{Geng:2025yys}
H.~Geng and L.~Randall,
{\it Holography and Causality in the Karch-Randall Braneworld},
[arXiv:2504.21856 [hep-th]].

\bibitem{Paul:2025gpk}
S.~Paul, G.~Guin and S.~Gangopadhyay,
{\it Holographic entanglement entropy and complexity for the cosmological braneworld model},
JHEP \textbf{08}, 164 (2025)
doi:10.1007/JHEP08(2025)164
[arXiv:2505.11553 [hep-th]].

\bibitem{Liu:2022bdu}
Y.~Liu, X.~J.~Wang, J.~P.~Wu and X.~Zhang,
``Holographic superfluid with gauge\textendash{}axion coupling,''
Eur. Phys. J. C \textbf{83}, no.8, 748 (2023)
doi:10.1140/epjc/s10052-023-11918-9
[arXiv:2212.01986 [hep-th]].




\bibitem{Nozaki:2012qd}
M.~Nozaki, T.~Takayanagi and T.~Ugajin,
{\it Central Charges for BCFTs and Holography},
JHEP \textbf{06}, 066 (2012)
doi:10.1007/JHEP06(2012)066
[arXiv:1205.1573 [hep-th]].

\bibitem{DosSantos:2022exb}
F.~F.~Dos Santos,
{\it Entanglement entropy in Horndeski gravity},
JHAP \textbf{3}, no.1, 1-14 (2022)
doi:10.22128/jhap.2022.488.1015
[arXiv:2201.02500 [hep-th]].

\bibitem{Ryu:2006bv}
S.~Ryu and T.~Takayanagi,
{\it Holographic derivation of entanglement entropy from AdS/CFT},
Phys. Rev. Lett. \textbf{96}, 181602 (2006)
doi:10.1103/PhysRevLett.96.181602
[arXiv:hep-th/0603001 [hep-th]].

\bibitem{Randall:1999vf}
L.~Randall and R.~Sundrum,
{\it An Alternative to compactification},
Phys. Rev. Lett. \textbf{83}, 4690-4693 (1999)
doi:10.1103/PhysRevLett.83.4690
[arXiv:hep-th/9906064 [hep-th]].


\bibitem{Siano}
Siano, F., and R. Egger. "Josephson current through a nanoscale magnetic quantum dot." Physical review letters 93.4 (2004): 047002, {doi.org/10.1103/PhysRevLett.93.047002}.

\bibitem{Chandrasekhar:2024epu}
V.~Chandrasekhar,
{\it Current dependence of the low bias resistance of small capacitance Josephson junctions},
Phys. Lett. A \textbf{507}, 129493 (2024)
doi:10.1016/j.physleta.2024.129493
[arXiv:2404.05890 [cond-mat.supr-con]].

\bibitem{Senapati}
Senapati, Tapas, Ashwin Kumar Karnad, and Kartik Senapati. "Phase biasing of a Josephson junction using Rashba–Edelstein effect." Nature Communications 14.1 (2023): 7415, {doi.org/10.1038/16204}.




\bibitem{A.Baselmans}
 J. J. A. Baselmans, A. F. Morpurgo, B. J. van Wees, and T. M. Klapwijk, Nature 397, 43-45
(1999), Reversing the direction of the supercurrent in a controllable Josephson junction, {doi.org/10.1038/16204}.



\bibitem{Erdmenger:2014xya}
J.~Erdmenger, M.~Flory and M.~N.~Newrzella,
{\it Bending branes for DCFT in two dimensions},
JHEP \textbf{01}, 058 (2015)
doi:10.1007/JHEP01(2015)058
[arXiv:1410.7811 [hep-th]].

\bibitem{Minamitsuji:2013vra}
M.~Minamitsuji,
{\it Braneworlds with field derivative coupling to the Einstein tensor},
Phys. Rev. D \textbf{89}, no.6, 064025 (2014)
doi:10.1103/PhysRevD.89.064025
[arXiv:1312.3760 [gr-qc]].

\bibitem{Kawamoto:2023wzj}
T.~Kawamoto, S.~M.~Ruan and T.~Takayanagi,
{\it Gluing AdS/CFT},
JHEP \textbf{07}, 080 (2023)
doi:10.1007/JHEP07(2023)080
[arXiv:2303.01247 [hep-th]].

\bibitem{Kanda:2023zse}
H.~Kanda, M.~Sato, Y.~k.~Suzuki, T.~Takayanagi and Z.~Wei,
{\it AdS/BCFT with brane-localized scalar field},
JHEP \textbf{03}, 105 (2023)
doi:10.1007/JHEP03(2023)105
[arXiv:2302.03895 [hep-th]].








\bibitem{Rinaldi:2012vy}
M.~Rinaldi,
{\it Black holes with non-minimal derivative coupling},
Phys. Rev. D \textbf{86} (2012), 084048
doi:10.1103/PhysRevD.86.084048
[arXiv:1208.0103 [gr-qc]].

\bibitem{Babichev:2013cya}
E.~Babichev and C.~Charmousis,
{\it Dressing a black hole with a time-dependent Galileon},
JHEP \textbf{08} (2014), 106
doi:10.1007/JHEP08(2014)106
[arXiv:1312.3204 [gr-qc]].

\bibitem{Anabalon:2013oea}
A.~Anabalon, A.~Cisterna and J.~Oliva,
{\it Asymptotically locally AdS and flat black holes in Horndeski theory},
Phys. Rev. D \textbf{89} (2014), 084050
doi:10.1103/PhysRevD.89.084050
[arXiv:1312.3597 [gr-qc]].

\bibitem{Hui:2012qt}
L.~Hui and A.~Nicolis,
{\it No-Hair Theorem for the Galileon},
Phys. Rev. Lett. \textbf{110} (2013), 241104
doi:10.1103/PhysRevLett.110.241104
[arXiv:1202.1296 [hep-th]].

\bibitem{Baggioli:2016rdj}
M.~Baggioli,
{\it Gravity, holography and applications to condensed matter},
[arXiv:1610.02681 [hep-th]].

\bibitem{Son:2002sd}
D.~T.~Son and A.~O.~Starinets,
{\it Minkowski space correlators in AdS / CFT correspondence: Recipe and applications},
JHEP \textbf{09}, 042 (2002)
doi:10.1088/1126-6708/2002/09/042
[arXiv:hep-th/0205051 [hep-th]].

\bibitem{Hartnoll:2009sz}
S.~A.~Hartnoll,
{\it Lectures on holographic methods for condensed matter physics},
Class. Quant. Grav. \textbf{26}, 224002 (2009)
doi:10.1088/0264-9381/26/22/224002
[arXiv:0903.3246 [hep-th]].



\bibitem{Jeon}
Jeon, K.-R. et al. Chiral antiferromagnetic josephson junctions as spin-triplet
supercurrent spin valves and dc squids. Nature Nanotechnology 1–7 (2023), {doi.org/10.1038/s41565-023-01336-z}.

\bibitem{Bhatia}
Bhatia, E. et al. Nanoscale domain wall engineered spin-triplet josephson junctions and squid. Nano Letters 21, 3092–3097 (2021), {pubs.acs.org/doi/abs/10.1021/acs.nanolett.1c00273}.

\bibitem{Robinson2007}
Robinson, J., Piano, S., Burnell, G., Bell, C. and Blamire, M. Zero to $\pi$ transition in superconductor-ferromagnet-superconductor junctions. Phys. Rev. B
76, 094522 (2007), {doi.org/10.1103/PhysRevB.76.094522}.

\bibitem{Robinson2006}
 Robinson, J., Piano, S., Burnell, G., Bell, C. and Blamire, M. Critical current oscilalations in strong ferromagnetic $\pi$ junctions. Phys. Rev. Lett. 97, 177003, (2006), {doi.org/10.1103/PhysRevLett.97.177003}.

\bibitem{Sickinger} 
Sickinger, H., et al. "Experimental evidence of a $\phi$ Josephson junction." Phys. Rev. Lett. 109.10 (2012): 107002, {
doi.org/10.1103/PhysRevLett.109.107002}.

\bibitem{Haxell:2023uci}
D.~Z.~Haxell, M.~Coraiola, M.~Hinderling, S.~C.~ten Kate, D.~Sabonis, A.~E.~Svetogorov, W.~Belzig, E.~Cheah, F.~Krizek and R.~Schott, \textit{et al.}
{\it Demonstration of nonlocal Josephson effect in Andreev molecules},
Nano Lett. \textbf{23}, 7532-7538 (2023)
doi:10.1021/acs.nanolett.3c02066
[arXiv:2306.00866 [cond-mat.supr-con]].

\bibitem{Paudel:2024xyy}
P.~P.~Paudel, N.~O.~Smith and T.~D.~Stanescu,
{\it Disorder effects in planar semiconductor-superconductor structures: Majorana wires versus Josephson junctions},
[arXiv:2405.12192 [cond-mat.supr-con]].

\bibitem{Hinderling:2024jfq}
M.~Hinderling, S.~C.~t.~Kate, M.~Coraiola, D.~Z.~Haxell, M.~Stiefel, M.~Mergenthaler, S.~Paredes, S.~W.~Bedell, D.~Sabonis and F.~Nichele,
{\it Direct Microwave Spectroscopy of Andreev Bound States in Planar Ge Josephson Junctions},
PRX Quantum \textbf{5}, no.3, 030357 (2024)
doi:10.1103/PRXQuantum.5.030357
[arXiv:2403.03800 [cond-mat.mes-hall]].

\bibitem{Marcos:2013aya}
D.~Marcos, P.~Rabl, E.~Rico and P.~Zoller,
{\it Superconducting Circuits for Quantum Simulation of Dynamical Gauge Fields},
Phys. Rev. Lett. \textbf{111}, no.11, 110504 (2013)
doi:10.1103/PhysRevLett.111.110504
[arXiv:1306.1674 [cond-mat.mes-hall]].


\bibitem{Simmonds:2004ucc}
R.~W.~Simmonds, K.~M.~Lang, D.~A.~Hite, S.~Nam, D.~P.~Pappas and J.~M.~Martinis,
{\it Decoherence in Josephson Phase Qubits from Junction Resonators},
Phys. Rev. Lett. \textbf{93}, no.7, 077003 (2004)
doi:10.1103/PhysRevLett.93.077003

\bibitem{Ruhtinas}
Ruhtinas, Aki, and Ilari J. Maasilta. "Highly tunable NbTiN Josephson junctions fabricated with focused helium ion beam. {arXiv:2303.17348 (2023)}.






\end{thebibliography}
\end{document}